\documentclass[aps,pra,reprint,groupedaddress]{revtex4-1}

\usepackage{bm, braket,amsmath,mathtools,comment}
\usepackage[dvipdfmx]{graphicx}
\usepackage{hyperref}
\begin{document}

%\preprint{}

\title{Cooling schemes for two-component fermions in layered optical lattices}


\author{Shimpei Goto}
\email[]{shimpei.goto@yukwa.kyoto-u.ac.jp}
%\thanks{}

\author{Ippei Danshita}
\affiliation{Yukawa Institute for Theoretical Physics, Kyoto University, Kitashirakawa Oiwakecho, Sakyo-ku, Kyoto 606-8502, Japan}

\date{\today}

\begin{abstract}
Recently, a cooling scheme for ultracold atoms in a bilayer optical lattice has been proposed [A. Kantian \textit{et al}., arXiv:1609.03579].
In their scheme, the energy offset between the two layers is increased dynamically such that the entropy of one layer is transferred to the other layer.
Using the full-Hilbert-space approach, we compute cooling dynamics subjected to the scheme in order to show that their scheme fails to cool down two-component fermions.
We develop an alternative cooling scheme for two-component fermions, 
in which the spin-exchange interaction of one layer is significantly reduced.
Using both full-Hilbert-space and matrix-product-state approaches, we find that our scheme can decrease the temperature of the other layer by roughly half.
\end{abstract}

%\keywords{}

\maketitle

\section{Introduction}
Thanks to their unprecedented controllability and cleanness, ultracold atom systems have been utilized as analog quantum simulators.
Such simulators allow one to tackle some important problems in quantum many-body physics that can not be addressed with classical computers because of the requirement of too large computational resources \cite{bloch_quantum_2012,georgescu_quantum_2014}.
Thus far, they have been successfully applied for revealing, e.g., thermodynamic \cite{horikoshi_measurement_2010,nascimbene_exploring_2010,ku_revealing_2012,horikoshi_ground-state_2016} and dynamical \cite{gaebler_observation_2010,yefsah_heavy_2013,husmann_connecting_2015} properties of strongly interacting two-component Fermi gases in the superfluid phase with s-wave pairing, 
Pomeranchuk cooling of the SU(6) Hubbard model \cite{taie_su6_2012}, 
far-from-equilibrium dynamics of the Hubbard model \cite{kondov_disorder-induced_2015,schreiber_observation_2015}, 
the Bose-Hubbard model \cite{greiner_collapse_2002,trotzky_probing_2012,choi_exploring_2016} and the Ising model \cite{meinert_quantum_2013}.

Quantum-simulation technology for the Hubbard model, which is naturally realized with two-component fermions in optical lattices \cite{mazurenko_cold-atom_2017,jordens_mott_2008,schneider_metallic_2008,greif_short-range_2013,hart_observation_2015,haller_single-atom_2015,edge_imaging_2015,parsons_site-resolved_2016,boll_spin-_2016,cheuk_observation_2016,brown_observation_2016}, has been highly demanded.
In the Hubbard system, it is expected that spin fluctuations mediate several nontrivial properties, such as frustrated magnetism \cite{starykh_unusual_2015,zhou_quantum_2017} and $d$-wave superconducting phase \cite{lee_doping_2006,maier_$mathitd$_2000}.
Hence, the Hubbard model is considered to be essential for understanding strong correlation effects on electrons in solids, especially, mechanisms of high-$T_\mathrm{c}$ superconductivity. 
However, accurate large-scale numerical simulations of this model are not feasible at present except in some limited spatial geometry or parameter regions, e.g., on a one-dimensional chain \cite{lieb_absence_1968} or at half filling \cite{staudt_phase_2000,otsuka_universal_2016}.

The most severe bottleneck for quantum simulations of the Hubbard model is to reduce the temperature of the systems to be much lower than the spin-exchange interaction $J = 4t^2 / U$ so that the above-mentioned interesting physics emerges. 
Here, $t$ and $U$ denote the hopping integral and onsite interaction of the Hubbard model.
For instance, the critical temperature of the $d$-wave superconducting phases is estimated to be on the order of $0.1J/k_\mathrm{B}$ with the use of the dynamical cluster approximation \cite{maier_$mathitd$_2000}.
Since the lowest temperature achieved in ultracold-atom quantum simulators is $0.45J/k_\mathrm{B}$ \cite{mazurenko_cold-atom_2017}, one needs to develop techniques for further cooling.

Recently, \textcite{kantian_dynamical_2016} have proposed a cooling scheme using a bilayer optical lattice. 
In this scheme, the entropy of one layer is transferred to the other layer in the following way.
Gases in both layers are initially prepared in a gapless phase, such as superfluid. 
The energy offset between the layers is adiabatically increased such that one of the layers takes a commensurate filling rate, at which the gas is in a gapped insulating phase, e.g., Mott insulator, while the other layer remains in the gapless phase.
Since the energy scale set by the excitation-energy gap in the insulating layer is much larger than that in the gapless layer, the entropy flows from the former to the latter.
In other words, the insulating layer is cooled down.
A remarkable advantage of this cooling scheme is that it can be combined with other cooling methods that have been used previously \cite{mckay_cooling_2011,onofrio_cooling_2017}, 
such as evaporative cooling \cite{masuhara_evaporative_1988,davis_bose-einstein_1995}, sympathetic cooling \cite{modugno_bose-einstein_2001,gunter_bose-fermi_2006}, and optimization of confinement potentials \cite{bernier_cooling_2009,mathy_enlarging_2012,hart_observation_2015,mazurenko_cold-atom_2017}.

However, it is still unclear whether or not the proposed scheme is effective for two-component fermions because quantitative numerical evaluation of this scheme is lacking. 
In contrast to the case of one-component bosons, the Mott insulator of two-component fermions, whose particle excitations are gapped, has gapless spin excitations such that the entropy does not necessarily flow from to the Mott-insulator layer to the other.
Since this cooling scheme may be potentially able to decrease experimentally available temperature down to the order of $0.1J/k_\mathrm{B}$, it is important to examine the validity and the performance of the scheme for the Hubbard systems with the use of accurate numerical methods.

In this paper, by means of the full-Hilbert-space (FHS) approach, we compute time evolution of the bilayer Hubbard model in one dimension at finite temperatures subjected to the dynamical parameter change corresponding to the cooling scheme of \textcite{kantian_dynamical_2016}.
We evaluate the performance of the scheme of Kantian \textit{et al}. for two-component fermions and indeed find it ineffective.
Alternatively, we propose a modified cooling scheme that is effective for two-component fermions in layered optical lattices.
In our scheme, the system is initially prepared to be the Mott insulator at half filling.
The spin-exchange interaction of one layer, which acts as a coolant, is adiabatically decreased such that the entropy of the other layer is absorbed to the coolant layer.
We show that the system can be cooled down to roughly half of the initial temperatures.
We also improve a way for describing a thermal mixed state at low temperatures within a matrix-product-state (MPS) approach and use it to confirm that our cooling scheme is effective for a system with larger size that is compatible with experiments.

The remainder of the paper is organized as follows.
In Sec.~\ref{sec:dynamical}, we describe the details of two dynamical cooling schemes simulated in this paper.
In Sec.~\ref{sec:numrical}, numerical methods used for simulations are explained. 
The improved MPS approach is also introduced in this section. 
Simulated data are shown in Sec.~\ref{sec:result}. 
We also discuss the performance of the cooling schemes deduced from the simulated data.
Conclusions are given in Sec.~\ref{sec:con}.

\section{\label{sec:dynamical}Model and dynamical processes for cooling}
In order to numerically examine the cooling schemes for two-component fermions in bilayer optical lattices, we specifically analyze the Hubbard model on a two-leg ladder lattice \cite{noack_correlations_1994,kuroki_quantum_1996,feiguin_pair_2009,yamamoto_trapped_2009}.
The Hamiltonian of this system at time $\tau$ is given by
\begin{align}
\hat{H}(\tau) = &\quad \hat{H}_A(\tau) + \hat{H}_B(\tau) + \hat{H}_\perp(\tau) \nonumber \\ &+ E(\tau) \sum^{N_r}_{i=1} \sum_{\sigma = \uparrow, \downarrow} \hat{n}_{Bi \sigma}, \\
\hat{H}_X(\tau)  = &-t_X(\tau) \sum^{N_r-1}_{i=1} \sum_{\sigma = \uparrow, \downarrow} (\hat{c}^\dagger_{Xi\sigma} \hat{c}_{Xi+1 \sigma} + \mathrm{H.c.}) \nonumber \\
&+ U_X(\tau) \sum^{N_r}_{i=1} \left( \hat{n}_{Xi \uparrow} - \frac{1}{2} \right) \left(\hat{n}_{Xi \downarrow} - \frac{1}{2}\right), \\
\hat{H}_\perp(\tau) = &-t_{\perp}(\tau) \sum^{N_r-1}_{i=1} \sum_{\sigma = \uparrow, \downarrow} (\hat{c}^\dagger_{Ai \sigma}\hat{c}_{Bi \sigma} + \mathrm{H.c.}).
\end{align}
Each layer consists of a 1D Hubbard chain whose Hamiltonian is denoted by $\hat{H}_X$.
Here, $X \in \{A, B\}$ represents the chain index, $N_r$ denotes the number of sites in each chain, $t_X(\tau)$ and $U_X(\tau)$ are the hopping integral and the onsite interaction in chain $X$, 
$E(\tau)$ is the energy offset between the two chains, 
$\hat{c}_{Xi\sigma}$ annihilates a fermion with spin $\sigma$ on site $i$ in chain $X$, and $\hat{n}_{Xi \sigma} = \hat{c}^\dagger_{Xi\sigma} \hat{c}_{Xi\sigma}$. 
The two chains are coupled by the interchain hopping $t_\perp(\tau)$.
At initial time $\tau = 0$, these parameters are set to be $t_X (0) = t_\perp (0) = t$, $U_X(0) = U$, and $E(0) = 0$.
We treat chains A and B as the target and coolant subsystems, respectively.
The optical lattice of two-leg ladder geometry can be created in experiments, e.g., by means of double-well optical lattice \cite{sebby-strabley_lattice_2006,folling_direct_2007,danshita_quantum_2007,chen_many-body_2011}.
Hereafter, we set $k_\mathrm{B} = \hbar = 1$ except in the figures and their captions.

We choose the specific two-leg ladder geometry because it allows for computing real-time dynamics of the quantum many-body system with accurate numerical methods, namely, FHS approaches for system with small size (we specifically takes $N_r = 4$) and MPS approaches for those with relatively large size ($N_r = 10$). 
Since the working mechanism of the analyzed cooling scheme does not rely on any specific properties of 1D systems, we believe that the conclusions drawn from our analysis should be applied to the Hubbard systems at higher dimensions at least qualitatively.

\begin{figure}
\includegraphics[scale=0.25]{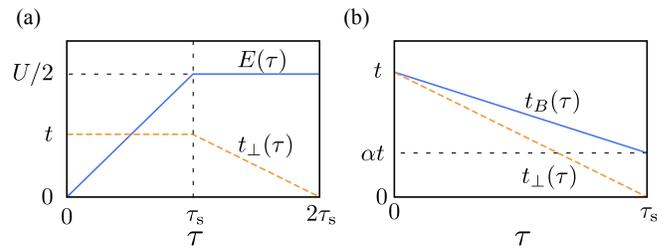}
\caption{\label{fig:protcol}
(Color online)
(a) The time sequences of the parameters varied in the EOV scheme. 
(a) The time sequences of the parameters varied in the SEV scheme. 
}
\end{figure}

\begin{figure}
\includegraphics[scale=0.25]{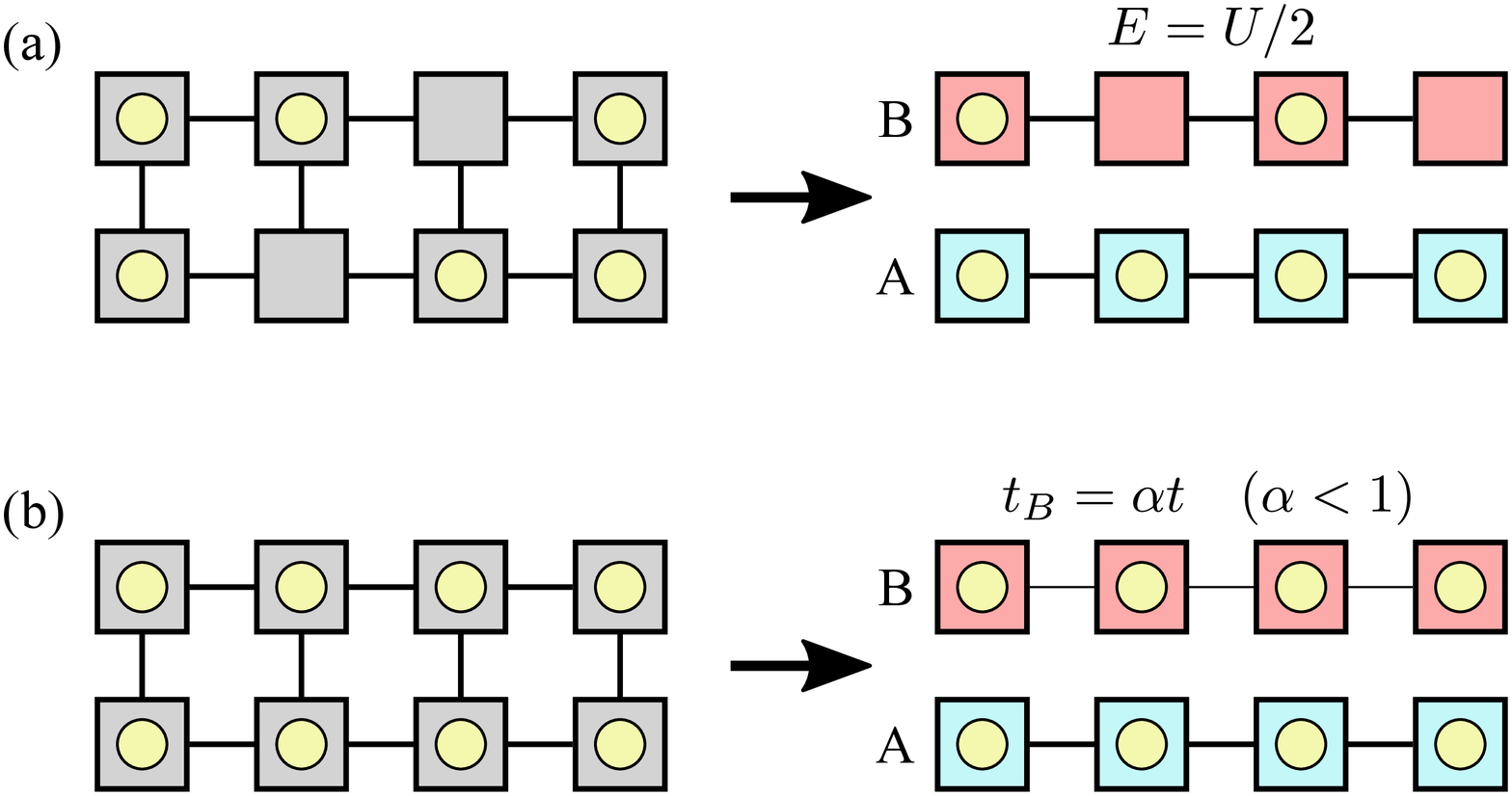}
\caption{\label{fig:schemes}
(Color online)
(a) Schematic picture of the EOV scheme. Initially, the filling of a system is lower than half filling. 
By increasing the energy offset up to $U/2$ and decreasing the interchain hopping, 
the system is separated to one metallic chain and one half-filled insulating chain. 
(b) Schematic picture of the SEV scheme. Initially, a system is in the Mott insulating state at half filling. 
By decreasing the hopping integral of chain B down to $\alpha t$ and also decreasing the interchain hopping to zero, 
the system is separated to two half-filled insulating chains. The energy scale of spin excitations in chain B is much smaller than that of chain A.}
\end{figure}

In the dynamical cooling scheme proposed in \textcite{kantian_dynamical_2016}, which we shall call the energy-offset-variation (EOV) scheme, the system is initially prepared in a gapless phase with delocalized particles such as superfluid or metal. 
One slowly increases the energy offset $E$ such that the target subsystem changes to a gapped insulating state while the coolant one remains in a gapless phase.
One next decreases the interchain hopping $t_\perp$ down to zero in order to isolate the target from the coolant.
If there is no remaining gapless excitation in the Mott insulating target, its local temperature significantly exceeds that in the gapless coolant such that the entropy flows from the target to the coolant.
In terms of thermodynamics, the increases of the energy offset can be interpreted as compression of the target in the sense that it leads to the increase of the energy scale set by low-lying excitations in the target.
\textcite{kantian_dynamical_2016} have confirmed that this scheme is effective for the one-component Bose-Hubbard system.

However, in the case of the two-component fermions, the required condition for the cooling scheme is unsatisfied, i.e., the spin excitations remain gapless in the Mott insulating phase at half filling .
Moreover, the bandwidth of the spin-excitation branch, which is on the order of the spin-exchange interaction $J$, is smaller than that of particle-excitation branch of the metallic phase so that the entropy may possibly flow from the coolant to the target in contrast to what we desire.

%[$f \equiv \sum^{2N_r}_i\braket{\hat{n}_{i\uparrow}}/(2N_r) = \sum^{2N_r}_i\braket{\hat{n}_{i\downarrow}}/(2N_r)$]
Since it is unclear at this stage whether or not the EOV scheme is effective for the Hubbard system, we examine the performance of this scheme on the basis of quantitative numerical calculations.
Specifically, we set the two-leg Hubbard system initially at $3/8$ filling, and each parameter is linearly varied as shown in Fig.~\ref{fig:protcol}(a), where $\tau_\mathrm{s}$ denotes the sweep time.
In this process, the initial metallic state on a two-leg ladder is separated into a half-filled Mott insulating chain and a metallic chain as depicted in Fig.~\ref{fig:schemes}(a).
At the end of this process, we estimate the temperature of the target subsystem from the internal energy and the static spin structure factor as discussed in Sec.~\ref{sec:thermo}.
We will indeed see that the EOV scheme is ineffective for the Hubbard system.

We propose an alternative cooling scheme for two-component fermions in a bilayer optical lattice, which we shall call the spin-exchange-variation (SEV) scheme.
In this scheme, we assume that the initial state is the Mott insulating state at half filling in the two-leg ladder such that the particle excitations are largely gapped and hardly involved in the cooling dynamics.
In the coolant subsystem (chain B), one slowly decreases the hopping integral $t_B$ or increases the onsite interaction $U_B$, implying that the spin-exchange interaction $J_B = 4 t^2_B/U_B$ is decreased.
The decrease of $J_B$ shrinks the bandwidth of the spin-excitation branch in the coolant.
In terms of thermodynamics, the decrease of the spin-exchange interaction can be interpreted as expansion of the coolant, which should lead to entropy flow from the target to the coolant. 
At the same time, one decreases the interchain hopping $t_\perp$ down to zero in order to isolate the target from the coolant.
The specific time sequences of $t_B(\tau)$ and $t_\perp(\tau)$ are illustrated in Fig.~\ref{fig:protcol}(b), where, $\alpha < 1$ is a parameter characterizing the reduction of the hopping $t_B(\tau)$.
After this process, the initial Mott insulating state on the two-leg ladder is separated into two independent Mott insulating chains as depicted in Fig.~\ref{fig:schemes}(b).
Notice that one does not change any parameters in chain A.

\section{\label{sec:numrical}Numerical methods}
In this section, we explain the two numerical methods, namely, the FHS approach and the MPS approach, which allow us to accurately compute real-time evolution of the density matrix of a quantum many-body system on a lattice.
We use the FHS approach for simulating both EOV and SEV schemes at a small system, say $N_r = 4$. 
Since the FHS calculations will eventually show that the SEV scheme is effective, we will double-check whether or not this is also the case for a larger system ($N_r = 10$) by means of the MPS approach. 
For this purpose, we make a little improvement on a MPS approach for representing a thermal mixed state at low temperature by changing the alignment of MPS via swap operations.

When $N_r = 4$, the dimensions $D$ of the full Hilbert space of the symmetry blocks that we are interested in are not so large: $D=3136$ for $(N_\uparrow, N_\downarrow) = (3, 3)$ sector corresponding to 3/8 filling and $D=4900$ for $(N_\uparrow, N_\downarrow) = (4, 4)$ sector corresponding to half filling. 
Here, $N_\sigma$ is the number of particles with spin $\sigma$.
Thus, we can explicitly construct the Hamiltonian matrix and take the matrix exponential of it to obtain a density matrix at temperature $T = 1 / \beta$ given by
\begin{equation}
\label{eq:rho}
\hat{\rho} = \frac{\exp(-\beta \hat{H})}{\mathrm{Tr}\exp(-\beta \hat{H})}. 
\end{equation}
The time evolution of the density matrix is described by the von-Neumann equation,
\begin{equation}
\label{eq:vNeq}
	\mathrm{i} \frac{\partial \hat{\rho}(\tau)}{\partial \tau} = \hat{H}(\tau)\hat{\rho}(\tau) - \hat{\rho}(\tau)\hat{H}(\tau).
\end{equation}
We numerically solve Eq.~\eqref{eq:vNeq} by using the fourth order Runge-Kutta method with the time step $\Delta \tau = 0.025 t^{-1}$.

For the larger system ($N_r = 10$), the dimensions of the symmetric subspaces are too large to perform the FHS calculation.
Hence, we use MPS \cite{schollwock_density-matrix_2011}, in which unnecessary states in the Hilbert space are efficiently truncated.
In the MPS representation, a state $\ket{\psi}$ in a $N$-site system is represented as 
\begin{equation} 
\label{eq:MPS}
\ket{\psi} = \sum_{\bm{\sigma}}\bm{A}^{\sigma_1}_1 \bm{A}^{\sigma_2}_2 \cdots \bm{A}^{\sigma_N}_N \ket{\bm{\sigma}},
\end{equation} 
where $\sigma_i$ is the state of the local Hilbert space at $i$-th site, e.g., $\sigma_i \in \{\ket{0}, \ket{\uparrow}, \ket{\downarrow}, \ket{\uparrow \downarrow}\}$ for two-component fermion systems,
$\ket{\bm{\sigma}} = \ket{\sigma_1, \sigma_2, \cdots, \sigma_N}$, and $\sum_{\bm{\sigma}}$ means the summation over all possible configurations of $\sigma_i$. 
The dimension of matrices $\bm{A}^{\sigma_i}_i$, which is often called bond dimension, needs to be exponentially large with respect to the system size $N$ for representing arbitrary states.
However, for certain classes of states, including ground states, low-lying excited states, and slow dynamics starting with them, the required bond dimension scales only polynomially with $N$, allowing for an efficient representation of these states with a MPS.

For representing a thermal mixed state given by Eq.~\eqref{eq:rho} within the MPS framework, there are major two options: a sampling approach \cite{white_minimally_2009,stoudenmire_minimally_2010} and an ancilla-site approach \cite{verstraete_matrix_2004,feiguin_finite-temperature_2005}.
In this work, we adopt the latter approach because we also needs to compute real-time evolution starting with a thermal mixed state and the latter approach has been shown to be more efficient for this purpose \cite{binder_minimally_2015}.
In the ancilla-site approach, one introduces ancilla sites, which are copies of physical sites consisting of all possible states with equal weight and are maximally entangled with them, to represent a mixed state at infinite temperature as a ``wave function'' $\ket{\psi_\infty}$.
Using $\ket{\psi_\infty}$, one can write down a MPS at temperature $T$ as \cite{feiguin_finite-temperature_2005}
\begin{equation}
\label{eq:T_MPS}
\ket{\psi_T} = \exp\left(-\frac{\beta}{2}\hat{H}\right) \ket{\psi_{\infty}}.
\end{equation}
Notice that the operator $\exp(-\beta \hat{H}/2)$ is applied only to matrices on the physical sites.
With this MPS, we can evaluate the thermal expectation value of an operator $\hat{O}$ as 
\begin{equation}
\label{eq:th_exp}
\braket{\hat{O}} = \frac{\braket{\psi_T|\hat{O}|\psi_T}}{\braket{\psi_T|\psi_T}}.
\end{equation}
In this work, we use the exact MPS representation to prepare the infinite temperature state of the canonical ensemble \cite{barthel_matrix_2016}.

In the ancilla-site approach, entanglement between the ancilla and physical states is interpreted as thermal fluctuations.
The price to pay for the inclusion of the thermal fluctuations is that the size of the entire Hilbert space is squared.
In return for the enlargement of the Hilbert space, we can treat a mixed state as a pure state such that all convenient techniques for MPS are available \cite{karrasch_finite-temperature_2012,karrasch_reducing_2013,tiegel_matrix_2014}.

We compute the thermal MPS of Eq.~\eqref{eq:T_MPS} via the imaginary-time evolution, in which $-\mathrm{i} \beta/2$ is interpreted as the propagation time.
The time evolution operator $\mathcal{U}(\tau + \Delta \tau, \tau)$ for small time step $\Delta \tau$ is approximated as 
\begin{align}
\label{eq:tevol}
\mathcal{U}(\tau + \Delta \tau, \tau)&= \mathcal{T}\exp \left[ -\mathrm{i} \int^{\tau + \Delta \tau}_\tau \hat{H}(s) \mathrm{d}s\right] \nonumber \\
                &\simeq \exp \left[ -\mathrm{i} \int^{\tau + \Delta \tau}_\tau \hat{H}(s) \mathrm{d}s\right] \nonumber \\
                & = \exp \left[ -\mathrm{i} \hat{H}\left(\tau+\frac{\Delta \tau}{2}\right)\Delta \tau \right],
\end{align}
where $\mathcal{T}$ denotes the time ordering operator and the last equality follows from the linear dependence of $\hat{H}(\tau)$ on $\tau$.
This approximation is equivalent to replacing the Hamiltonian within a short time span $[\tau, \tau+\Delta \tau]$ with $H(\tau + \Delta \tau / 2)$.
The application of the time evolution operator on a MPS is implemented with the truncated Taylor expansion up to 10th order (See Appendix \ref{sec:Taylor} for details).
Notice that the Hamiltonian is independent of time in the imaginary time evolution while Eq.~\eqref{eq:tevol} is expressed in a more general form that is applicable to the Hamiltonian dependent linearly on time.

\begin{figure}
\includegraphics[scale=0.3]{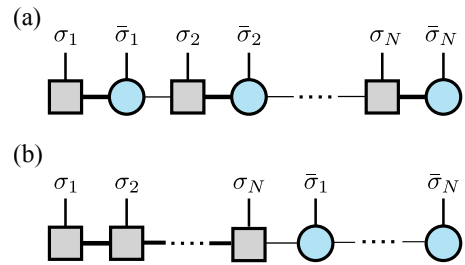}
\caption{\label{fig:alignment} 
(Color online)
(a) The alignment of alternating physical and ancilla sites.
Squares and circles represent physical and ancilla sites. 
A vertical line represents a physical index $\sigma_i$ or a copy of it for ancilla site $\bar{\sigma}_i$.
(b) The efficient alignment for low temperatures.
}
\end{figure}

The efficiency of a MPS representation depends strongly on how we align sites in the MPS representation.
Specifically, if two strongly entangled sites are more distant from each other in the MPS representation, it requires the bond dimension to be larger. 
In this sense, the MPS at infinite temperature is efficiently represented by the alignment of alternating physical and ancilla sites as shown in Fig.~\ref{fig:alignment}(a) \cite{schollwock_density-matrix_2011}, 
because each local physical state is maximally entangled with the corresponding ancilla state.
When temperature decreases, thermal fluctuations become weaker so that entanglement between the physical and ancilla sites becomes weaker as well.
In contrast, the entanglement among physical sites grows.
At zero temperature, for instance, thermal fluctuations are absent so that only the entanglement among the physical sites is important.
This means that the alignment of sites shown in Fig.~\ref{fig:alignment}(a) unnecessarily increases the required bond dimension at low temperature.
A recent study, in which the sampling approach and the ancilla-site approach are compared \cite{bruognolo_matrix_2017}, has indeed reported that the ancilla-site approach is less efficient in a low temperature region.

We overcome this inefficiency at low temperature by rearranging the MPS in the form illustrated in Fig.~\ref{fig:alignment}(b), where all the physical (ancilla) sites are assembled on the left (right) side of the MPS.
With this alignment, the required bond dimension for a thermal mixed state is comparable to that for the ground state without ancilla sites at least near zero temperature, i.e., the ancilla-site approach is efficient for low temperature systems.
The detailed procedure for the rearrangement is given in Appendix \ref{sec:rearrange}.
With such an approach, one practical question arises: when do we rearrange the MPS?
In this work, we choose an extreme option: we do it immediately after we operate the first time-evolution operator on the infinite-temperature MPS.
This option is not optimized for describing high-temperature systems but suited for describing low-temperature systems, which we are interested in.

\begin{figure}
\includegraphics[scale=0.4]{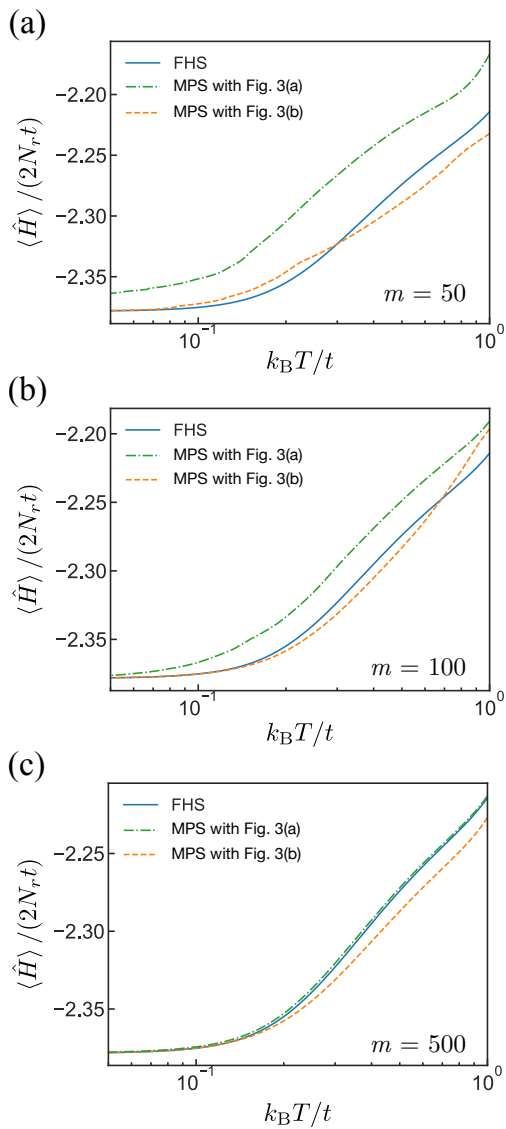}
\caption{\label{fig:mdep} 
(Color online) 
The internal energy $\braket{\hat{H}}$ versus the temperature $T$ at $\tau = 0$, where $N_r = 4$, $(N_\uparrow, N_\downarrow) = (4, 4)$, and $U/t = 8.0$.
The blue solid, green dash-dot, and orange dashed lines represent the results obtained by the FHS approach, the MPS with the alignment of Fig.~\ref{fig:alignment}(a), and the MPS with the alignment of Fig.~\ref{fig:alignment}(b).
In MPS calculations, we set the time step for the imaginary time evolution to be $\Delta \tau = - \mathrm{i}0.05 \hbar/t$ and the maximum bond dimension to be $m=50$ (a), $100$ (b), and $500$ (c).
}
\end{figure}

In order to corroborate that the MPS approach with the alignment of Fig.~\ref{fig:alignment}(b) is more efficient at low temperatures, we compare the temperature dependence of the internal energy obtained with the two alignments in Fig.~\ref{fig:mdep}. 
We also use several maximum bond dimensions of the MPS $m$ to see convergence properties of these approaches. 
At sufficiently large bond dimension, e.g., $m=500$, the internal energies computed by the MPS approach with the alignment of Fig.~\ref{fig:alignment}(a) (green dash-dot line) agree well with those given by the FHS approach (blue solid line) in the entire temperature region. 
However, the agreement is rather poor for relatively small bond dimensions ($m=50$, $100$). 
In contrast, the internal energies computed by the MPS approach with the alignment of Fig.~\ref{fig:alignment}(b) (orange dashed line) agree with the FHS results in a low temperature region even at the small bond dimensions.

\begin{figure}
\includegraphics[scale=0.4]{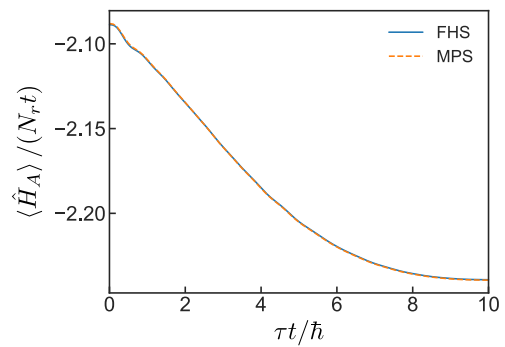}
\caption{\label{fig:comp} 
(Color online) 
The time evolution of the internal energy of chain A $\braket{\hat{H}_A}$ during the SEV scheme computed by the FHS (blue solid line) and MPS (orange dashed line) approaches, where $N_r = 4$, $(N_\uparrow, N_\downarrow) = (4, 4)$, $U/t = 8.0$, $\tau_\mathrm{s} t = 10.0$, $\alpha = 0.1$, and $k_\mathrm{B}T/t = 0.25$.
We set the maximum bond dimension for the MPS to be $m = 1000$.
The time step $\Delta \tau$ is $0.1\hbar/t$ for the MPS approach and $0.05\hbar/t$ for the FHS approach.
}
\end{figure}

In this work, in addition to thermal equilibrium states at finite temperature, we compute real-time evolution of these states subjected to the dynamical variation of the Hamiltonian mentioned in Sec.~\ref{sec:dynamical}.
In Fig.~\ref{fig:comp}, we simulate the SEV scheme and show the real-time evolution of the internal energies computed by the FHS approach (blue solid line) and the MPS approach (orange dashed line).
There we choose $T/t = 0.25$, which is the lowest temperature that has been achieved experimentally in ultracold fermions in an optical lattice \cite{mazurenko_cold-atom_2017}. 
There is no discernible difference between the two results, corroborating that the MPS approach accurately captures the real-time dynamics subjected to the cooling scheme.

\section{\label{sec:result}Performance of dynamical cooling schemes in fermion systems}
\subsection{\label{sec:thermo}Thermometers}
\begin{figure}
\includegraphics[scale=0.27]{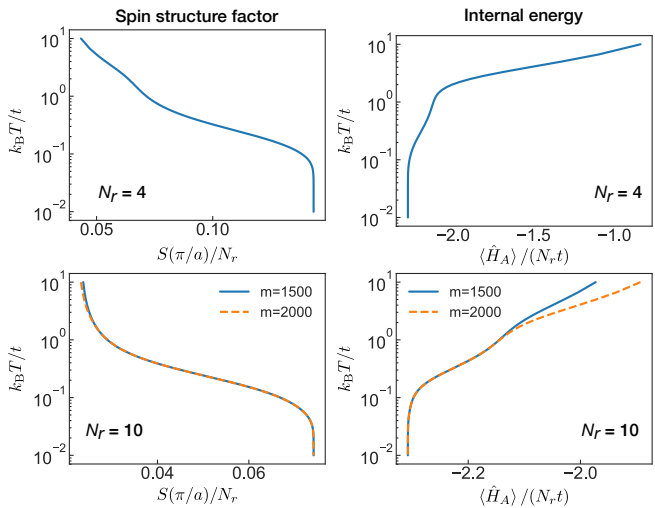}
\caption{\label{fig:thermo} 
(Color online) 
Numerical data used as thermometers in this work. For $N_r=4$, the data are obtained by the FHS approach. For $N_r = 10$, whose data are obtained by the MPS approach, 
we calculate physical quantities twice with different maximum bond dimensions $m$. We can use these data as thermometers in the region $k_{\mathrm{B}}T/t < 1.0$ where the data for two different $m$ converge.}
\end{figure}

In order to judge whether or not the target subsystem is actually cooled after the cooling scheme, we need to measure the temperature of the target, which is not an observable that can be calculated through Eq.~\eqref{eq:th_exp}.
In order to estimate the temperature, we calculate the internal energy and the static spin structure factor $S(k)$ at $k = \pi / a$ as functions of the temperature for thermal equilibrium states of the target subsystem (chain A) at half filling in the symmetric sector $N_\uparrow + N_\downarrow = N_r$.
Here, $S(k)$ is given by
\begin{equation}
S(k) = \frac{1}{N_r}\sum_{ij} \braket{\hat{S}^z_i \hat{S}^z_j}\mathrm{e}^{-\mathrm{i}k(r_i-r_j)}
\end{equation}
where $\hat{S}^z_i = (\hat{n}_{i \uparrow} - \hat{n}_{i \downarrow})/2$, $r_i = ia$, and $a$ is the lattice spacing.
As shown in Fig. \ref{fig:thermo}, both quantities are monotonic functions of $T$ so that the temperature is uniquely determined from each quantity.

\subsection{Comparison of two schemes}
\begin{figure}
\includegraphics[scale=0.45]{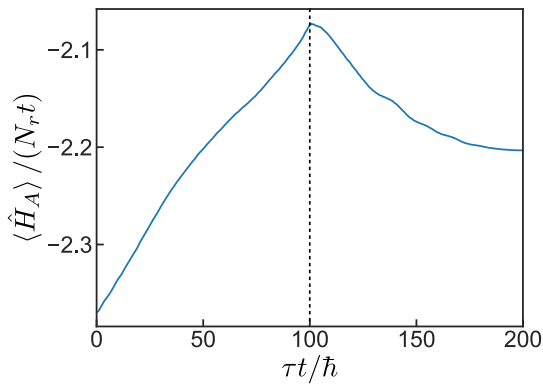}
\caption{\label{fig:met-ins} 
(Color online) 
The time evolution of the internal energy of chain A during the EOV cooling scheme with the sweep time $\tau_\mathrm{s} t/ \hbar = 100$, where $N_r =4$, $(N_\uparrow, N_\downarrow) = (3, 3)$, and $U/t = 8.0$.
We increase the energy offset linearly during the sweep time $\tau_\mathrm{s}$ and next decrease the interchain hopping linearly during the same sweep time.
Temperature after the cooling estimated from the internal energy at $\tau t/\hbar = 200$ is $0.29t/k_\mathrm{B}$, which is slightly higher than the initial temperature $0.25t/k_\mathrm{B}$.}
\end{figure}
We first examine the performance of the EOV scheme by using the FHS approach for $N_r = 4$.
Figure \ref{fig:met-ins} represents the time evolution of the internal energy of chain A (target of the cooling) during the dynamical process based on the EOV scheme with the sweep time $\tau_\mathrm{s}t = 100$.
After the cooling process, namely $\tau = 2 \tau_\mathrm{s}$, $\braket{\hat{H}_A}/(N_r t) = -2.203$ and $S(\pi/a)/N_r = 0.102$.
The temperature estimated from the internal energy is $T_{\mathrm{est}}/t \approx 0.29$ and that from the spin structure factor is $T_{\mathrm{est}}/t \approx 0.31$.
Both estimated temperatures are higher than the initial temperature $T/t = 0.25$.
This result clearly shows that chain A is heated up rather than cooled down with the EOV scheme.
we also confirm the transfer of the entropy from chain B to chain A by comparing the von Neumann entropy from the reduced density matrix.
Thus, the EOV scheme is not effective for cooling of two-component fermions.
As discussed in Sec.~\ref{sec:dynamical}, this failure of the EOV scheme can be attributed to the presence of gapless spin excitations in the Mott insulating state.

\begin{figure}
\includegraphics[scale=0.45]{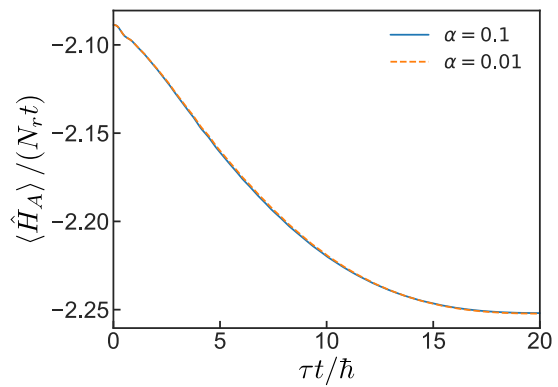}
\caption{\label{fig:spin-scheme} 
(Color online) 
The time evolution of the internal energy of chain A during the SEV scheme with $\alpha = 0.1$ and $\alpha = 0.01$, where $N_r = 4$, $(N_\uparrow, N_\downarrow) = (4, 4)$, $U/t = 8.0$, and $\tau_\mathrm{s} t/ \hbar = 20$. 
From the value of the internal energy after the time evolution ($\tau = \tau_\mathrm{s}$), we estimate the cooled temperature of chain A as $0.15t/k_\mathrm{B}$, which is roughly half of the initial temperature $0.25t/k_\mathrm{B}$.}
\end{figure}

We next simulate the SEV scheme with use of the FHS approach for $N_r = 4$.
Figure \ref{fig:spin-scheme} represents the time evolution of $\braket{\hat{H}_A}$ under the SEV scheme with $\tau_\mathrm{s} t = 20$ and $\alpha = 0.1$.
After the cooling process, namely at $\tau = \tau_\mathrm{s}$, $\braket{\hat{H}_A}/(N_r t) = -2.252$ and $S(\pi/a)/N_r = 0.131$, which correspond to $T_{\mathrm{est}}/t \approx 0.15$ and $T_{\mathrm{est}}/t \approx 0.14$, respectively.
We recall that the initial temperature is $T/t = 0.25$.
Thus, the SEV scheme can significantly reduce the temperature even in the presence of gapless spin excitations.
The transfer of the entropy from chain A to chain B is also confirmed.
We also perform the simulation with $\alpha = 0.01$ and confirm that further decrease in $\alpha$ makes little differences as shown in Fig.~\ref{fig:spin-scheme}.

\begin{figure}
\includegraphics[scale=0.45]{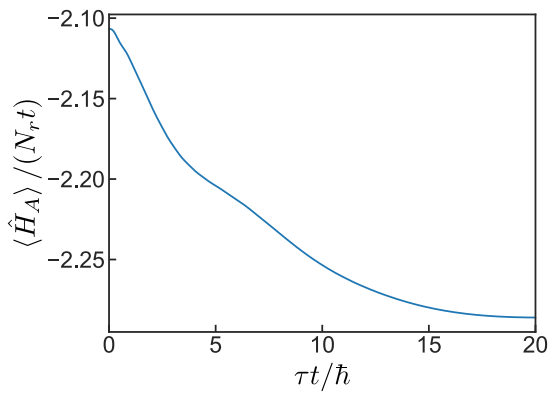}
\caption{\label{fig:larger} 
(Color online) 
The time evolution of the internal energy of chain A with $\alpha = 0.1$, where $N_r = 10$, $(N_\uparrow, N_\downarrow) = (10, 10)$, $U/t = 8.0$, $\tau_\mathrm{s}t/\hbar = 20$, and $k_\mathrm{B}T/t = 0.25$. 
The estimated temperature from the internal energy is $0.14t/k_\mathrm{B}$.}
\end{figure}
In order to check that the SEV scheme is effective also for a system with large size, we set $N_r = 10$ and apply the MPS approach for simulating the SEV scheme with the same $\tau_\mathrm{s} t = 20$ and $\alpha = 0.1$.
The resulting time evolution of $\braket{\hat{H}_A}$ is shown in Fig.~\ref{fig:larger}.
The estimated temperatures after the cooling are $T_\mathrm{est}/t \approx 0.13$ from the spin structure factor and $T_\mathrm{est}/t \approx 0.14$ from the internal energy.
Although the $N_r=10$ system is more than twice as large as the $N_r = 4$ system, the SEV scheme with the same sweep time exhibits almost the same performance for both systems. 
Therefore, the sweep time required to perform the cooling has little dependency on the size of systems so that this cooling scheme is expected to be effective in much larger systems.

Comparing the simulated data of the EOV and the SEV schemes, we conclude that the latter scheme is more suited for cooling the Hubbard system than the former.
Furthermore, the SEV is expected to be effective regardless of the system size.
In the following subsection, we investigate the performance of this cooling scheme in greater details.

\subsection{Further performance test of the spin-exchange-variation scheme}
Since we have confirmed that the SEV scheme at $N_r = 10$ gives almost the same performance as that at $N_r=4$, we further examine the performance of the SEV scheme with the FHS simulations in the smaller system.

\begin{figure}
\includegraphics[scale=0.45]{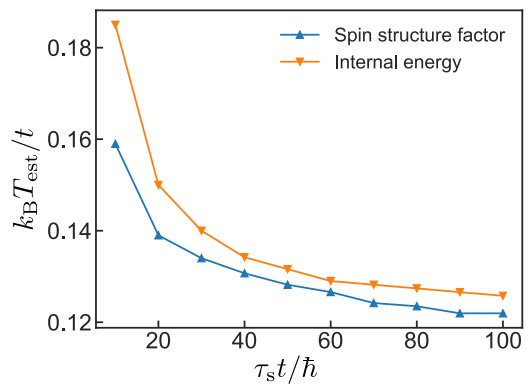}
\caption{\label{fig:TauDep} 
(Color online) 
The sweep time dependence of the estimated temperatures from the internal energy and the spin structure factor, where $N_r = 4$, $(N_\uparrow, N_\downarrow) = (4, 4)$, $U/t = 8.0$, and $k_\mathrm{B}T/t = 0.25$.
}
\end{figure}
Figure \ref{fig:TauDep} shows the sweep time dependences of the estimated temperatures.
For $\tau_\mathrm{s} t > 50$, the estimated temperatures almost converge to certain values, which are roughly half of the initial temperature $T/t = 0.25$.
Since the energy scale of the hopping integral $t$ is on the order of 1 kHz in experiments \cite{mazurenko_cold-atom_2017}, 
the sweep time $50 t^{-1}$ is on the order of 10 ms.
With this timescale, the SEV scheme can halve the temperature of the Hubbard system. 

\begin{figure}
\includegraphics[scale=0.45]{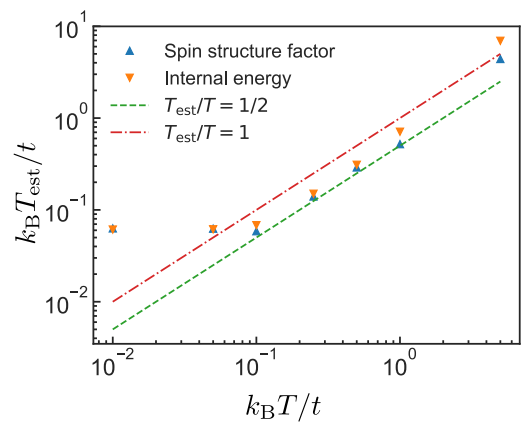}
\caption{\label{fig:Tdep} 
(Color online) 
The estimated temperatures from the internal energy and the spin structure factor as functions of the initial temperature, where $N_r = 4$, $(N_\uparrow, N_\downarrow) = (4, 4)$, $U/t = 8.0$, and the sweep time $\tau_\mathrm{s} t = 20$.}
\end{figure}
As explained in Sec.~\ref{sec:dynamical}, the working mechanism of the SEV scheme is based on the assumption that in the initial Mott insulator the particle excitations are largely gapped so that the relevant degrees of freedom in the cooling dynamics are the gapless spin excitations.
Let us check what will happen if the assumption is broken by increasing the initial temperature to be comparable or above the energy gap of the particle excitations.
Figure \ref{fig:Tdep} shows the initial temperature dependences of the estimated temperatures after the SEV cooling scheme with the sweep time $\tau_\mathrm{s} t = 20$.
The SEV scheme is effective in the temperature regions $0.1 \leq T/t \leq 1.0$ from the obtained data.
For high initial temperatures, e.g., $T/t = 5.0$, which is comparable to $U/t$, excitations other than spin ones are thermally induced such that the assumption for the cooling scheme is violated.
As expected, there the SEV scheme is ineffective.
In this sense, the SEV scheme is useful for the further cooling of a system already cooled by other cooling schemes.

Figure \ref{fig:Tdep} indicates that the SEV scheme is also ineffective when low initial temperature is as low as $T/t \leq 0.05$. 
This ineffectiveness comes from too fast sweep of parameters.
In other words, the inverse of the used sweep time $\tau_\mathrm{s} t = 20$ is comparable to or larger than the initial temperature,
 and such a nonadiabatic dynamical process heats up the system.
In order to corroborate this interpretation, we perform a numerical simulation with larger sweep time $\tau_\mathrm{s} t = 50$ at the initial temperature $T/t = 0.05$. 
The estimated temperature from the internal energy decreases down to $T_\mathrm{est}/t \approx 0.033$. 
From the above observations, it is expected that the SEV scheme with a sweep time $\tau_\mathrm{s}$ can cool the system down to a temperature whose order is $1/(\tau_\mathrm{s} t)$.

\begin{figure}
\includegraphics[scale=0.45]{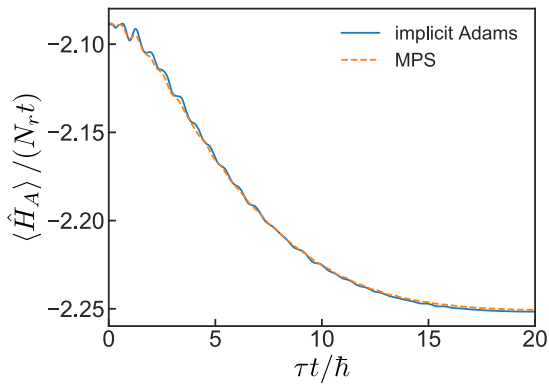}
\caption{\label{fig:u_scheme} 
(Color online) 
The time evolution of the internal energy of chain A during the SEV scheme with the increase of $U_B$, where $N_r = 4$, $(N_\uparrow, N_\downarrow) = (4, 4)$, $U/t = 8.0$, and $\tau_\mathrm{s}t = 20$. We increase $U_B$ linearly from $U_B/t = 8.0$ to $80.0$.
The blue solid and orange dashed lines represent the data computed by the FHS approach with the implicit Adams method and the MPS approaches with $\Delta \tau = 0.01 \hbar/t$.}
\end{figure}
In the numerical simulations of the SEV scheme performed so far, we slowly decrease the hopping integral in chain B $t_B$ to decrease $J_B=4t_B^2/U_B$.
However, the SEV scheme can be accomplished by increasing the Hubbard interaction $U_B$ in chain B.
In specific, we increase $U_B$ linearly from $8.0t$ up to $80.0t$ instead of decreasing $t_B$.
Figure \ref{fig:u_scheme} shows the time evolutions of the internal energy during the SEV scheme implemented with the increase of $U_B $with the sweep time $\tau_\mathrm{s} t = 20$.
In this simulation, the fourth order Runge-Kutta method is numerically unstable due to the large $U_B$. 
Instead, we adopt the implicit Adams method \cite{brown_vode:_1989} and the MPS method with the 10-th order truncated Taylor expansion, which are more numerically stable. 
The simulated data from the implicit Adams method are good agreement with those from the MPS method.
Thus, there is no numerical instability in the simulations.
Estimations from the spin structure factor and the internal energy give $T_\mathrm{est}/t \approx 0.15$.
This performance of the cooling is as almost the same as that of the scheme implemented by decreasing the hopping integral with the same sweep time.

In the viewpoint of the accessibility in experiments, the SEV scheme with the increase of $U_B$ is preferable because one can control the onsite interaction in a spatially dependent way by means of state-of-art optical techniques, such as the optical Feshbach resonance \cite{fedichev_influence_1996,yamazaki_submicron_2010} and the optically induced magnetic Feshbach resonance \cite{clark_quantum_2015}. 
On the other hand, the spatial control of the hopping integral required for the SEV cooling is relatively more challenging and has not been realized although such control is possible in principle by using the optical-lattice microscope techniques with single-site resolution.

We finally note an advantage of the SEV scheme that it can be successively repeated $l$ times if one starts with a system with $2^l$ layers.
Let us assume that the system is initially at $T/J=0.45$, which is the lowest temperature that has ever been achieved experimentally in two-component fermions \cite{mazurenko_cold-atom_2017}.
We also anticipate that a single SEV cooling roughly halves the temperature.
Performing the SEV cooling twice will bring the system to a low temperature on the order of $T/J=0.1$, where many interesting properties resulting from spin fluctuations are expected to emerge.

\section{\label{sec:con}Conclusions}
We proposed a cooling scheme for two-component fermions in layered optical lattices, which we call the spin-exchange-variation (SEV) scheme.
The SEV scheme may be viewed as a modified version of the cooling scheme proposed by \textcite{kantian_dynamical_2016}, which we call the energy-offset-variation (EOV) scheme.
From the exact numerical simulations based on the full-Hilbert-space (FHS) approach, we confirmed that the EOV scheme cannot cool two-component fermions because of the presence of gapless spin excitations in the Mott insulating state.
Using the matrix-product-state approach in addition to the FHS approach, we showed that the EOV scheme is so effective that it can decrease the temperature down to the roughly half of its initial value.
The cooling scheme using layered geometry opens up new possibilities for ultracold-atom quantum simulators to access physics of the Hubbard model governed by spin fluctuations, such as high-$T_\mathrm{c}$ superconductivity and frustrated magnetism.

\begin{acknowledgments}
The authors thank Y. Takahashi, Y. Takasu, and G. Watanabe for useful comments and discussions.
The MPS calculations in this work are performed using the ITensor library, http://itensor.org.
This work was financially supported by KAKENHI from Japan Society for the Promotion of Science: Grant No. 25220711 and CREST, JST No. JPMJCR1673.
\end{acknowledgments}

\appendix
\section{\label{sec:Taylor}Implementation of the truncated Taylor expansion}
Applying the exponential of an operator $\hat{O}$ to a state $\ket{\psi}$ is a common task in quantum mechanics. 
Here, we consider the situation where the state is given by a MPS form (\ref{eq:MPS}) and the operator $\hat{O}$ is given by a matrix product operator (MPO) form
\begin{align}
\label{eq:MPO}
\hat{O} = \sum_{\bm{\sigma}\bm{\sigma}^\prime} 
\bm{W}^{\sigma_1, \sigma^\prime_1}_1 \bm{W}^{\sigma_2, \sigma^\prime_2}_2 \cdots \bm{W}^{\sigma_N, \sigma^\prime_N}_N \ket{\bm{\sigma}}\bra{\bm{\sigma}^\prime},
\end{align}
where $\bm{W}^{\sigma_i, \sigma^\prime_i}_i$ is a matrix whose coefficient is operators acting on the local Hilbert space at site $i$.
Within the framework of the MPS and the MPO, there are variant approaches to achieve this task \cite{garcia-ripoll_time_2006}.
In this work, we adopt an approach which is simple, easy to implement, and applicable to Hamiltonians with long-range interactions: the Taylor expansion.

In this approach, the exponential of an operator is approximated by the truncated Taylor expansion up to $M$-th order as
\begin{align}
\label{eq:Taylor}
\exp(\hat{O}) \ket{\psi} \simeq \left( \sum^M_{n=0} \frac{\hat{O}^n}{n!}\right)\ket{\psi}.
\end{align}
For the $M$-th order expansion, $M$ times additions of MPSs and $M$ times multiplications of the MPO to the MPS are required according to the Horner's method.
A crucial point for an efficient implementation is that one should avoid these additions, because the addition of MPSs generally increases the bond dimension of a MPS and the extra compressions of MPSs are required \cite{schollwock_density-matrix_2011}.
To avoid the additions, we factorize the truncated Taylor expansion (\ref{eq:Taylor}) as
\begin{align}
\label{eq:prod_exp}
\left( \sum^M_{n=0} \frac{\hat{O}^n}{n!}\right)\ket{\psi} = \left [ \prod^M_{i=0}(\openone - z_i\hat{O}) \right] \ket{\psi}.
\end{align}
The fourth order expansion is given in Ref.~\cite{garcia-ripoll_time_2006}.
In this appendix, we present an explicit way to construct the MPO for $\openone - z_i \hat{O}$ and to obtain the expansion Eq.~\eqref{eq:prod_exp} up to arbitrary order.

The operation $(\openone - z_i \hat{O})$ can be represented by an MPO whose matrix dimension is the same as that of the MPO of $\hat{O}$.
In general, $\bm{W}^{\sigma_i, \sigma^\prime_{i}}_i$ whose matrix dimension is $d_i \times d_{i+1}$ can be represented by block triangular form \cite{zaletel_time-evolving_2015} with a $1 \times (d_{i+1} -2)$ matrix $\bm{X}_i$, a $1 \times 1$ matrix $\bm{Y}_i$, a $(d_i-2) \times (d_{i+1} - 2)$ matrix $\bm{U}_i$, and a $(d_{i}-2 \times 1)$ matrix $\bm{V}_i$ as 
\begin{widetext}
\begin{align}
\label{eq:GenMPO}
\bm{W}^{\sigma_1, \sigma^\prime_1}_1 \bm{W}^{\sigma_2, \sigma^\prime_2}_2 \cdots \bm{W}^{\sigma_N, \sigma^\prime_N}_N = 
\left (
\begin{array}{ccc}
\openone & \bm{X}_1 & \bm{Y}_1
\end{array}
\right )
\left (
\begin{array}{ccc}
\openone & \bm{X}_2 & \bm{Y}_2 \\
0 & \bm{U}_2 & \bm{V}_2 \\
0 & 0 & \openone 
\end{array}
\right )
\cdots
\left (
\begin{array}{c}
\bm{Y}_N \\ \bm{V}_N \\ \openone
\end{array}
\right ).
\end{align}
\end{widetext}
For this representation, an MPO for $(\openone - z_i \hat{O})$ is given by 
replacing $\bm{W}^{\sigma_1, \sigma^\prime_1}_1$ with 
%\begin{widetext}
\begin{equation}
\bm{W}^{\sigma_1, \sigma^\prime_1}_1  = 
\left (
\begin{array}{ccc}
-z_i \openone & -z_i\bm{X}_1 & -z_i \bm{Y}_1 + \openone
\end{array}
\right )
.
\end{equation}
%\end{widetext}
As shown, we can obtain the MPO for $(\openone - z_i \hat{O})$ by only modifying $\bm{W}^{\sigma_1, \sigma^\prime_1}_1$ and its bond dimensions are the same as those of the MPO $\hat{O}$.

The remaining task is to determine the factors $z_i$.
To obtain the factors $z_i$, we expand a polynomial 
\begin{align}
P(x) = \sum^M_{n=0}\frac{1}{n!} x^n
\end{align}
as
\begin{align}
\label{eq:poly_expand}
P(x) = \prod^M_{i=1}\left(1- \frac{x}{r_i}\right).
\end{align}
Here, $r_i$ is $i$-th root of the polynomial $P(x)$.
Since the identity operator $\openone$ and any operator $\hat{O}$ commute, the same expansion as Eq.~\eqref{eq:poly_expand} is valid when one replaces $1$ and $x$ with $\openone$ and $\hat{O}$, respectively.
From the comparison between Eqs.~\eqref{eq:prod_exp} and \eqref{eq:poly_expand}, the factors $z_i$ are given by the inverse of the roots $r_i$ and complex in general.
Consequently, we can implement the $M$-th order truncated Taylor expansion only by $M$ times multiplications of MPOs to a MPS.

\section{\label{sec:rearrange} 
How to rearrange the alignment of matrix product states}
\begin{figure}
\centering
\includegraphics[scale=0.2]{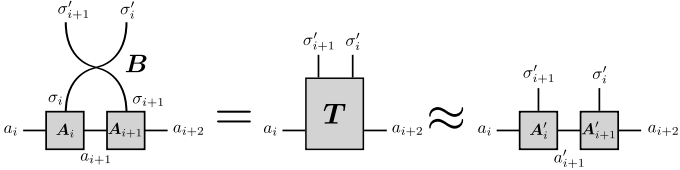}
\caption{\label{fig:swap} 
(Color online) 
The diagrammatic representation of Eq. (\ref{eq:swap}).}

\end{figure}

In this appendix, we show how a MPS is rearranged from the alignment in Fig. \ref{fig:alignment}(a) to that in Fig.~\ref{fig:alignment}(b).
For neighboring sites in a MPS representation, one can swap their positions by using the fermionic swap gate \cite{corboz_fermionic_2009}.
The fermionic swap gate is a rank-four tensor which is defined as 
\begin{align}
B^{\sigma^\prime_{i+1}, \sigma^\prime_i}_{\sigma_i, \sigma_{i+1}} = \delta_{\sigma^\prime_i, \sigma_i} \delta_{\sigma^\prime_{i+1}, \sigma_{i+1}} S[P(\sigma_i), P(\sigma_{i+1})],
\end{align}
with
\begin{align}
S[P(\sigma_i), P(\sigma_{i+1})] = 1 - 2 \delta_{P(\sigma_i), -1} \delta_{P(\sigma_{i+1}), -1}
\end{align}
where $\delta_{i, j}$ is the Kronecker's delta, $P(\sigma_i)$ is the fermion parity of the state $\sigma_i$, i.e., 
$P(\ket{\uparrow}) = P(\ket{\downarrow}) = -1$ and $P(\ket{0}) = P(\ket{\uparrow\downarrow}) = 1$.
As depicted in Fig.~\ref{fig:swap}, one can swap the positions of indices $\sigma_i$ and $\sigma_{i+1}$ by applying the swap gate.
After the application of the swap gate, the local matrices at sites $i$ and $i+1$ form a rank-four tensor.
One can decompose this rank-four tensor with the singular value decomposition to two rank-three tensors as  
\begin{align}
\label{eq:swap}
B^{\sigma^\prime_{i+1}, \sigma^\prime_i}_{\sigma_i, \sigma_{i+1}} A^{\sigma_i}_{a_i, a_{i+1}} A^{\sigma_{i+1}}_{a_{i+1}, a_{i+2}}
& = T^{\sigma^\prime_{i+1}, \sigma^\prime_i}_{a_i, a_{i+2}} \nonumber \\
& \approx A^{\prime \sigma^\prime_{i+1}}_{a_i, a^\prime_{i+1}} A^{\prime \sigma^\prime_i}_{a^\prime_{i+1}, a_{i+2}}.
\end{align}
Here, repeated indices are summed over.
Consequently, the positions of neighboring sites in a MPS representation are swapped.
The sequence of swaps to produce a required alignment can be obtained by sorting algorithms implemented with only neighboring swaps such as the bubble sort. 
%\begin{acknowledgments}
%\end{acknowledgments}

%\input{main.bbl}
\bibliographystyle{apsrev4-1}
\bibliography{library}

\begin{thebibliography}{69}%
\makeatletter
\providecommand \@ifxundefined [1]{%
 \@ifx{#1\undefined}
}%
\providecommand \@ifnum [1]{%
 \ifnum #1\expandafter \@firstoftwo
 \else \expandafter \@secondoftwo
 \fi
}%
\providecommand \@ifx [1]{%
 \ifx #1\expandafter \@firstoftwo
 \else \expandafter \@secondoftwo
 \fi
}%
\providecommand \natexlab [1]{#1}%
\providecommand \enquote  [1]{``#1''}%
\providecommand \bibnamefont  [1]{#1}%
\providecommand \bibfnamefont [1]{#1}%
\providecommand \citenamefont [1]{#1}%
\providecommand \href@noop [0]{\@secondoftwo}%
\providecommand \href [0]{\begingroup \@sanitize@url \@href}%
\providecommand \@href[1]{\@@startlink{#1}\@@href}%
\providecommand \@@href[1]{\endgroup#1\@@endlink}%
\providecommand \@sanitize@url [0]{\catcode `\\12\catcode `\$12\catcode
  `\&12\catcode `\#12\catcode `\^12\catcode `\_12\catcode `\%12\relax}%
\providecommand \@@startlink[1]{}%
\providecommand \@@endlink[0]{}%
\providecommand \url  [0]{\begingroup\@sanitize@url \@url }%
\providecommand \@url [1]{\endgroup\@href {#1}{\urlprefix }}%
\providecommand \urlprefix  [0]{URL }%
\providecommand \Eprint [0]{\href }%
\providecommand \doibase [0]{http://dx.doi.org/}%
\providecommand \selectlanguage [0]{\@gobble}%
\providecommand \bibinfo  [0]{\@secondoftwo}%
\providecommand \bibfield  [0]{\@secondoftwo}%
\providecommand \translation [1]{[#1]}%
\providecommand \BibitemOpen [0]{}%
\providecommand \bibitemStop [0]{}%
\providecommand \bibitemNoStop [0]{.\EOS\space}%
\providecommand \EOS [0]{\spacefactor3000\relax}%
\providecommand \BibitemShut  [1]{\csname bibitem#1\endcsname}%
\let\auto@bib@innerbib\@empty
%</preamble>
\bibitem [{\citenamefont {Bloch}\ \emph {et~al.}(2012)\citenamefont {Bloch},
  \citenamefont {Dalibard},\ and\ \citenamefont
  {Nascimb{\`e}ne}}]{bloch_quantum_2012}%
  \BibitemOpen
  \bibfield  {author} {\bibinfo {author} {\bibfnamefont {I.}~\bibnamefont
  {Bloch}}, \bibinfo {author} {\bibfnamefont {J.}~\bibnamefont {Dalibard}}, \
  and\ \bibinfo {author} {\bibfnamefont {S.}~\bibnamefont {Nascimb{\`e}ne}},\
  }\href {\doibase 10.1038/nphys2259} {\bibfield  {journal} {\bibinfo
  {journal} {Nat Phys}\ }\textbf {\bibinfo {volume} {8}},\ \bibinfo {pages}
  {267} (\bibinfo {year} {2012})}\BibitemShut {NoStop}%
\bibitem [{\citenamefont {Georgescu}\ \emph {et~al.}(2014)\citenamefont
  {Georgescu}, \citenamefont {Ashhab},\ and\ \citenamefont
  {Nori}}]{georgescu_quantum_2014}%
  \BibitemOpen
  \bibfield  {author} {\bibinfo {author} {\bibfnamefont {I.~M.}\ \bibnamefont
  {Georgescu}}, \bibinfo {author} {\bibfnamefont {S.}~\bibnamefont {Ashhab}}, \
  and\ \bibinfo {author} {\bibfnamefont {F.}~\bibnamefont {Nori}},\ }\href
  {\doibase 10.1103/RevModPhys.86.153} {\bibfield  {journal} {\bibinfo
  {journal} {Rev. Mod. Phys.}\ }\textbf {\bibinfo {volume} {86}},\ \bibinfo
  {pages} {153} (\bibinfo {year} {2014})}\BibitemShut {NoStop}%
\bibitem [{\citenamefont {Horikoshi}\ \emph {et~al.}(2010)\citenamefont
  {Horikoshi}, \citenamefont {Nakajima}, \citenamefont {Ueda},\ and\
  \citenamefont {Mukaiyama}}]{horikoshi_measurement_2010}%
  \BibitemOpen
  \bibfield  {author} {\bibinfo {author} {\bibfnamefont {M.}~\bibnamefont
  {Horikoshi}}, \bibinfo {author} {\bibfnamefont {S.}~\bibnamefont {Nakajima}},
  \bibinfo {author} {\bibfnamefont {M.}~\bibnamefont {Ueda}}, \ and\ \bibinfo
  {author} {\bibfnamefont {T.}~\bibnamefont {Mukaiyama}},\ }\href {\doibase
  10.1126/science.1183012} {\bibfield  {journal} {\bibinfo  {journal}
  {Science}\ }\textbf {\bibinfo {volume} {327}},\ \bibinfo {pages} {442}
  (\bibinfo {year} {2010})}\BibitemShut {NoStop}%
\bibitem [{\citenamefont {Nascimb{\`e}ne}\ \emph {et~al.}(2010)\citenamefont
  {Nascimb{\`e}ne}, \citenamefont {Navon}, \citenamefont {Jiang}, \citenamefont
  {Chevy},\ and\ \citenamefont {Salomon}}]{nascimbene_exploring_2010}%
  \BibitemOpen
  \bibfield  {author} {\bibinfo {author} {\bibfnamefont {S.}~\bibnamefont
  {Nascimb{\`e}ne}}, \bibinfo {author} {\bibfnamefont {N.}~\bibnamefont
  {Navon}}, \bibinfo {author} {\bibfnamefont {K.~J.}\ \bibnamefont {Jiang}},
  \bibinfo {author} {\bibfnamefont {F.}~\bibnamefont {Chevy}}, \ and\ \bibinfo
  {author} {\bibfnamefont {C.}~\bibnamefont {Salomon}},\ }\href {\doibase
  10.1038/nature08814} {\bibfield  {journal} {\bibinfo  {journal} {Nature}\
  }\textbf {\bibinfo {volume} {463}},\ \bibinfo {pages} {1057} (\bibinfo {year}
  {2010})}\BibitemShut {NoStop}%
\bibitem [{\citenamefont {Ku}\ \emph {et~al.}(2012)\citenamefont {Ku},
  \citenamefont {Sommer}, \citenamefont {Cheuk},\ and\ \citenamefont
  {Zwierlein}}]{ku_revealing_2012}%
  \BibitemOpen
  \bibfield  {author} {\bibinfo {author} {\bibfnamefont {M.~J.~H.}\
  \bibnamefont {Ku}}, \bibinfo {author} {\bibfnamefont {A.~T.}\ \bibnamefont
  {Sommer}}, \bibinfo {author} {\bibfnamefont {L.~W.}\ \bibnamefont {Cheuk}}, \
  and\ \bibinfo {author} {\bibfnamefont {M.~W.}\ \bibnamefont {Zwierlein}},\
  }\href {\doibase 10.1126/science.1214987} {\bibfield  {journal} {\bibinfo
  {journal} {Science}\ }\textbf {\bibinfo {volume} {335}},\ \bibinfo {pages}
  {563} (\bibinfo {year} {2012})}\BibitemShut {NoStop}%
\bibitem [{\citenamefont {Horikoshi}\ \emph {et~al.}(2016)\citenamefont
  {Horikoshi}, \citenamefont {Koashi}, \citenamefont {Tajima}, \citenamefont
  {Ohashi},\ and\ \citenamefont
  {Kuwata-Gonokami}}]{horikoshi_ground-state_2016}%
  \BibitemOpen
  \bibfield  {author} {\bibinfo {author} {\bibfnamefont {M.}~\bibnamefont
  {Horikoshi}}, \bibinfo {author} {\bibfnamefont {M.}~\bibnamefont {Koashi}},
  \bibinfo {author} {\bibfnamefont {H.}~\bibnamefont {Tajima}}, \bibinfo
  {author} {\bibfnamefont {Y.}~\bibnamefont {Ohashi}}, \ and\ \bibinfo {author}
  {\bibfnamefont {M.}~\bibnamefont {Kuwata-Gonokami}},\ }\href
  {http://arxiv.org/abs/1612.04026} {\bibfield  {journal} {\bibinfo  {journal}
  {arXiv:1612.04026 [cond-mat]}\ } (\bibinfo {year} {2016})},\ \bibinfo {note}
  {arXiv: 1612.04026}\BibitemShut {NoStop}%
\bibitem [{\citenamefont {Gaebler}\ \emph {et~al.}(2010)\citenamefont
  {Gaebler}, \citenamefont {Stewart}, \citenamefont {Drake}, \citenamefont
  {Jin}, \citenamefont {Perali}, \citenamefont {Pieri},\ and\ \citenamefont
  {Strinati}}]{gaebler_observation_2010}%
  \BibitemOpen
  \bibfield  {author} {\bibinfo {author} {\bibfnamefont {J.~P.}\ \bibnamefont
  {Gaebler}}, \bibinfo {author} {\bibfnamefont {J.~T.}\ \bibnamefont
  {Stewart}}, \bibinfo {author} {\bibfnamefont {T.~E.}\ \bibnamefont {Drake}},
  \bibinfo {author} {\bibfnamefont {D.~S.}\ \bibnamefont {Jin}}, \bibinfo
  {author} {\bibfnamefont {A.}~\bibnamefont {Perali}}, \bibinfo {author}
  {\bibfnamefont {P.}~\bibnamefont {Pieri}}, \ and\ \bibinfo {author}
  {\bibfnamefont {G.~C.}\ \bibnamefont {Strinati}},\ }\href {\doibase
  10.1038/nphys1709} {\bibfield  {journal} {\bibinfo  {journal} {Nat Phys}\
  }\textbf {\bibinfo {volume} {6}},\ \bibinfo {pages} {569} (\bibinfo {year}
  {2010})}\BibitemShut {NoStop}%
\bibitem [{\citenamefont {Yefsah}\ \emph {et~al.}(2013)\citenamefont {Yefsah},
  \citenamefont {Sommer}, \citenamefont {Ku}, \citenamefont {Cheuk},
  \citenamefont {Ji}, \citenamefont {Bakr},\ and\ \citenamefont
  {Zwierlein}}]{yefsah_heavy_2013}%
  \BibitemOpen
  \bibfield  {author} {\bibinfo {author} {\bibfnamefont {T.}~\bibnamefont
  {Yefsah}}, \bibinfo {author} {\bibfnamefont {A.~T.}\ \bibnamefont {Sommer}},
  \bibinfo {author} {\bibfnamefont {M.~J.~H.}\ \bibnamefont {Ku}}, \bibinfo
  {author} {\bibfnamefont {L.~W.}\ \bibnamefont {Cheuk}}, \bibinfo {author}
  {\bibfnamefont {W.}~\bibnamefont {Ji}}, \bibinfo {author} {\bibfnamefont
  {W.~S.}\ \bibnamefont {Bakr}}, \ and\ \bibinfo {author} {\bibfnamefont
  {M.~W.}\ \bibnamefont {Zwierlein}},\ }\href {\doibase 10.1038/nature12338}
  {\bibfield  {journal} {\bibinfo  {journal} {Nature}\ }\textbf {\bibinfo
  {volume} {499}},\ \bibinfo {pages} {426} (\bibinfo {year}
  {2013})}\BibitemShut {NoStop}%
\bibitem [{\citenamefont {Husmann}\ \emph {et~al.}(2015)\citenamefont
  {Husmann}, \citenamefont {Uchino}, \citenamefont {Krinner}, \citenamefont
  {Lebrat}, \citenamefont {Giamarchi}, \citenamefont {Esslinger},\ and\
  \citenamefont {Brantut}}]{husmann_connecting_2015}%
  \BibitemOpen
  \bibfield  {author} {\bibinfo {author} {\bibfnamefont {D.}~\bibnamefont
  {Husmann}}, \bibinfo {author} {\bibfnamefont {S.}~\bibnamefont {Uchino}},
  \bibinfo {author} {\bibfnamefont {S.}~\bibnamefont {Krinner}}, \bibinfo
  {author} {\bibfnamefont {M.}~\bibnamefont {Lebrat}}, \bibinfo {author}
  {\bibfnamefont {T.}~\bibnamefont {Giamarchi}}, \bibinfo {author}
  {\bibfnamefont {T.}~\bibnamefont {Esslinger}}, \ and\ \bibinfo {author}
  {\bibfnamefont {J.-P.}\ \bibnamefont {Brantut}},\ }\href {\doibase
  10.1126/science.aac9584} {\bibfield  {journal} {\bibinfo  {journal}
  {Science}\ }\textbf {\bibinfo {volume} {350}},\ \bibinfo {pages} {1498}
  (\bibinfo {year} {2015})}\BibitemShut {NoStop}%
\bibitem [{\citenamefont {Taie}\ \emph {et~al.}(2012)\citenamefont {Taie},
  \citenamefont {Yamazaki}, \citenamefont {Sugawa},\ and\ \citenamefont
  {Takahashi}}]{taie_su6_2012}%
  \BibitemOpen
  \bibfield  {author} {\bibinfo {author} {\bibfnamefont {S.}~\bibnamefont
  {Taie}}, \bibinfo {author} {\bibfnamefont {R.}~\bibnamefont {Yamazaki}},
  \bibinfo {author} {\bibfnamefont {S.}~\bibnamefont {Sugawa}}, \ and\ \bibinfo
  {author} {\bibfnamefont {Y.}~\bibnamefont {Takahashi}},\ }\href {\doibase
  10.1038/nphys2430} {\bibfield  {journal} {\bibinfo  {journal} {Nat Phys}\
  }\textbf {\bibinfo {volume} {8}},\ \bibinfo {pages} {825} (\bibinfo {year}
  {2012})}\BibitemShut {NoStop}%
\bibitem [{\citenamefont {Kondov}\ \emph {et~al.}(2015)\citenamefont {Kondov},
  \citenamefont {McGehee}, \citenamefont {Xu},\ and\ \citenamefont
  {DeMarco}}]{kondov_disorder-induced_2015}%
  \BibitemOpen
  \bibfield  {author} {\bibinfo {author} {\bibfnamefont {S.~S.}\ \bibnamefont
  {Kondov}}, \bibinfo {author} {\bibfnamefont {W.~R.}\ \bibnamefont {McGehee}},
  \bibinfo {author} {\bibfnamefont {W.}~\bibnamefont {Xu}}, \ and\ \bibinfo
  {author} {\bibfnamefont {B.}~\bibnamefont {DeMarco}},\ }\href {\doibase
  10.1103/PhysRevLett.114.083002} {\bibfield  {journal} {\bibinfo  {journal}
  {Phys. Rev. Lett.}\ }\textbf {\bibinfo {volume} {114}},\ \bibinfo {pages}
  {083002} (\bibinfo {year} {2015})}\BibitemShut {NoStop}%
\bibitem [{\citenamefont {Schreiber}\ \emph {et~al.}(2015)\citenamefont
  {Schreiber}, \citenamefont {Hodgman}, \citenamefont {Bordia}, \citenamefont
  {L{\"u}schen}, \citenamefont {Fischer}, \citenamefont {Vosk}, \citenamefont
  {Altman}, \citenamefont {Schneider},\ and\ \citenamefont
  {Bloch}}]{schreiber_observation_2015}%
  \BibitemOpen
  \bibfield  {author} {\bibinfo {author} {\bibfnamefont {M.}~\bibnamefont
  {Schreiber}}, \bibinfo {author} {\bibfnamefont {S.~S.}\ \bibnamefont
  {Hodgman}}, \bibinfo {author} {\bibfnamefont {P.}~\bibnamefont {Bordia}},
  \bibinfo {author} {\bibfnamefont {H.~P.}\ \bibnamefont {L{\"u}schen}},
  \bibinfo {author} {\bibfnamefont {M.~H.}\ \bibnamefont {Fischer}}, \bibinfo
  {author} {\bibfnamefont {R.}~\bibnamefont {Vosk}}, \bibinfo {author}
  {\bibfnamefont {E.}~\bibnamefont {Altman}}, \bibinfo {author} {\bibfnamefont
  {U.}~\bibnamefont {Schneider}}, \ and\ \bibinfo {author} {\bibfnamefont
  {I.}~\bibnamefont {Bloch}},\ }\href {\doibase 10.1126/science.aaa7432}
  {\bibfield  {journal} {\bibinfo  {journal} {Science}\ }\textbf {\bibinfo
  {volume} {349}},\ \bibinfo {pages} {842} (\bibinfo {year}
  {2015})}\BibitemShut {NoStop}%
\bibitem [{\citenamefont {Greiner}\ \emph {et~al.}(2002)\citenamefont
  {Greiner}, \citenamefont {Mandel}, \citenamefont {H{\"a}nsch},\ and\
  \citenamefont {Bloch}}]{greiner_collapse_2002}%
  \BibitemOpen
  \bibfield  {author} {\bibinfo {author} {\bibfnamefont {M.}~\bibnamefont
  {Greiner}}, \bibinfo {author} {\bibfnamefont {O.}~\bibnamefont {Mandel}},
  \bibinfo {author} {\bibfnamefont {T.~W.}\ \bibnamefont {H{\"a}nsch}}, \ and\
  \bibinfo {author} {\bibfnamefont {I.}~\bibnamefont {Bloch}},\ }\href
  {\doibase 10.1038/nature00968} {\bibfield  {journal} {\bibinfo  {journal}
  {Nature}\ }\textbf {\bibinfo {volume} {419}},\ \bibinfo {pages} {51}
  (\bibinfo {year} {2002})}\BibitemShut {NoStop}%
\bibitem [{\citenamefont {Trotzky}\ \emph {et~al.}(2012)\citenamefont
  {Trotzky}, \citenamefont {Chen}, \citenamefont {Flesch}, \citenamefont
  {McCulloch}, \citenamefont {Schollw{\"o}ck}, \citenamefont {Eisert},\ and\
  \citenamefont {Bloch}}]{trotzky_probing_2012}%
  \BibitemOpen
  \bibfield  {author} {\bibinfo {author} {\bibfnamefont {S.}~\bibnamefont
  {Trotzky}}, \bibinfo {author} {\bibfnamefont {Y.-A.}\ \bibnamefont {Chen}},
  \bibinfo {author} {\bibfnamefont {A.}~\bibnamefont {Flesch}}, \bibinfo
  {author} {\bibfnamefont {I.~P.}\ \bibnamefont {McCulloch}}, \bibinfo {author}
  {\bibfnamefont {U.}~\bibnamefont {Schollw{\"o}ck}}, \bibinfo {author}
  {\bibfnamefont {J.}~\bibnamefont {Eisert}}, \ and\ \bibinfo {author}
  {\bibfnamefont {I.}~\bibnamefont {Bloch}},\ }\href {\doibase
  10.1038/nphys2232} {\bibfield  {journal} {\bibinfo  {journal} {Nat Phys}\
  }\textbf {\bibinfo {volume} {8}},\ \bibinfo {pages} {325} (\bibinfo {year}
  {2012})}\BibitemShut {NoStop}%
\bibitem [{\citenamefont {Choi}\ \emph {et~al.}(2016)\citenamefont {Choi},
  \citenamefont {Hild}, \citenamefont {Zeiher}, \citenamefont {Schau{\ss}},
  \citenamefont {Rubio-Abadal}, \citenamefont {Yefsah}, \citenamefont
  {Khemani}, \citenamefont {Huse}, \citenamefont {Bloch},\ and\ \citenamefont
  {Gross}}]{choi_exploring_2016}%
  \BibitemOpen
  \bibfield  {author} {\bibinfo {author} {\bibfnamefont {J.-y.}\ \bibnamefont
  {Choi}}, \bibinfo {author} {\bibfnamefont {S.}~\bibnamefont {Hild}}, \bibinfo
  {author} {\bibfnamefont {J.}~\bibnamefont {Zeiher}}, \bibinfo {author}
  {\bibfnamefont {P.}~\bibnamefont {Schau{\ss}}}, \bibinfo {author}
  {\bibfnamefont {A.}~\bibnamefont {Rubio-Abadal}}, \bibinfo {author}
  {\bibfnamefont {T.}~\bibnamefont {Yefsah}}, \bibinfo {author} {\bibfnamefont
  {V.}~\bibnamefont {Khemani}}, \bibinfo {author} {\bibfnamefont {D.~A.}\
  \bibnamefont {Huse}}, \bibinfo {author} {\bibfnamefont {I.}~\bibnamefont
  {Bloch}}, \ and\ \bibinfo {author} {\bibfnamefont {C.}~\bibnamefont
  {Gross}},\ }\href {\doibase 10.1126/science.aaf8834} {\bibfield  {journal}
  {\bibinfo  {journal} {Science}\ }\textbf {\bibinfo {volume} {352}},\ \bibinfo
  {pages} {1547} (\bibinfo {year} {2016})}\BibitemShut {NoStop}%
\bibitem [{\citenamefont {Meinert}\ \emph {et~al.}(2013)\citenamefont
  {Meinert}, \citenamefont {Mark}, \citenamefont {Kirilov}, \citenamefont
  {Lauber}, \citenamefont {Weinmann}, \citenamefont {Daley},\ and\
  \citenamefont {N{\"a}gerl}}]{meinert_quantum_2013}%
  \BibitemOpen
  \bibfield  {author} {\bibinfo {author} {\bibfnamefont {F.}~\bibnamefont
  {Meinert}}, \bibinfo {author} {\bibfnamefont {M.~J.}\ \bibnamefont {Mark}},
  \bibinfo {author} {\bibfnamefont {E.}~\bibnamefont {Kirilov}}, \bibinfo
  {author} {\bibfnamefont {K.}~\bibnamefont {Lauber}}, \bibinfo {author}
  {\bibfnamefont {P.}~\bibnamefont {Weinmann}}, \bibinfo {author}
  {\bibfnamefont {A.~J.}\ \bibnamefont {Daley}}, \ and\ \bibinfo {author}
  {\bibfnamefont {H.-C.}\ \bibnamefont {N{\"a}gerl}},\ }\href {\doibase
  10.1103/PhysRevLett.111.053003} {\bibfield  {journal} {\bibinfo  {journal}
  {Phys. Rev. Lett.}\ }\textbf {\bibinfo {volume} {111}},\ \bibinfo {pages}
  {053003} (\bibinfo {year} {2013})}\BibitemShut {NoStop}%
\bibitem [{\citenamefont {Mazurenko}\ \emph {et~al.}(2017)\citenamefont
  {Mazurenko}, \citenamefont {Chiu}, \citenamefont {Ji}, \citenamefont
  {Parsons}, \citenamefont {Kan{\'a}sz-Nagy}, \citenamefont {Schmidt},
  \citenamefont {Grusdt}, \citenamefont {Demler}, \citenamefont {Greif},\ and\
  \citenamefont {Greiner}}]{mazurenko_cold-atom_2017}%
  \BibitemOpen
  \bibfield  {author} {\bibinfo {author} {\bibfnamefont {A.}~\bibnamefont
  {Mazurenko}}, \bibinfo {author} {\bibfnamefont {C.~S.}\ \bibnamefont {Chiu}},
  \bibinfo {author} {\bibfnamefont {G.}~\bibnamefont {Ji}}, \bibinfo {author}
  {\bibfnamefont {M.~F.}\ \bibnamefont {Parsons}}, \bibinfo {author}
  {\bibfnamefont {M.}~\bibnamefont {Kan{\'a}sz-Nagy}}, \bibinfo {author}
  {\bibfnamefont {R.}~\bibnamefont {Schmidt}}, \bibinfo {author} {\bibfnamefont
  {F.}~\bibnamefont {Grusdt}}, \bibinfo {author} {\bibfnamefont
  {E.}~\bibnamefont {Demler}}, \bibinfo {author} {\bibfnamefont
  {D.}~\bibnamefont {Greif}}, \ and\ \bibinfo {author} {\bibfnamefont
  {M.}~\bibnamefont {Greiner}},\ }\href {\doibase 10.1038/nature22362}
  {\bibfield  {journal} {\bibinfo  {journal} {Nature}\ }\textbf {\bibinfo
  {volume} {545}},\ \bibinfo {pages} {462} (\bibinfo {year}
  {2017})}\BibitemShut {NoStop}%
\bibitem [{\citenamefont {J{\"o}rdens}\ \emph {et~al.}(2008)\citenamefont
  {J{\"o}rdens}, \citenamefont {Strohmaier}, \citenamefont {G{\"u}nter},
  \citenamefont {Moritz},\ and\ \citenamefont {Esslinger}}]{jordens_mott_2008}%
  \BibitemOpen
  \bibfield  {author} {\bibinfo {author} {\bibfnamefont {R.}~\bibnamefont
  {J{\"o}rdens}}, \bibinfo {author} {\bibfnamefont {N.}~\bibnamefont
  {Strohmaier}}, \bibinfo {author} {\bibfnamefont {K.}~\bibnamefont
  {G{\"u}nter}}, \bibinfo {author} {\bibfnamefont {H.}~\bibnamefont {Moritz}},
  \ and\ \bibinfo {author} {\bibfnamefont {T.}~\bibnamefont {Esslinger}},\
  }\href {\doibase 10.1038/nature07244} {\bibfield  {journal} {\bibinfo
  {journal} {Nature}\ }\textbf {\bibinfo {volume} {455}},\ \bibinfo {pages}
  {204} (\bibinfo {year} {2008})}\BibitemShut {NoStop}%
\bibitem [{\citenamefont {Schneider}\ \emph {et~al.}(2008)\citenamefont
  {Schneider}, \citenamefont {Hackerm{\"u}ller}, \citenamefont {Will},
  \citenamefont {Best}, \citenamefont {Bloch}, \citenamefont {Costi},
  \citenamefont {Helmes}, \citenamefont {Rasch},\ and\ \citenamefont
  {Rosch}}]{schneider_metallic_2008}%
  \BibitemOpen
  \bibfield  {author} {\bibinfo {author} {\bibfnamefont {U.}~\bibnamefont
  {Schneider}}, \bibinfo {author} {\bibfnamefont {L.}~\bibnamefont
  {Hackerm{\"u}ller}}, \bibinfo {author} {\bibfnamefont {S.}~\bibnamefont
  {Will}}, \bibinfo {author} {\bibfnamefont {T.}~\bibnamefont {Best}}, \bibinfo
  {author} {\bibfnamefont {I.}~\bibnamefont {Bloch}}, \bibinfo {author}
  {\bibfnamefont {T.~A.}\ \bibnamefont {Costi}}, \bibinfo {author}
  {\bibfnamefont {R.~W.}\ \bibnamefont {Helmes}}, \bibinfo {author}
  {\bibfnamefont {D.}~\bibnamefont {Rasch}}, \ and\ \bibinfo {author}
  {\bibfnamefont {A.}~\bibnamefont {Rosch}},\ }\href {\doibase
  10.1126/science.1165449} {\bibfield  {journal} {\bibinfo  {journal}
  {Science}\ }\textbf {\bibinfo {volume} {322}},\ \bibinfo {pages} {1520}
  (\bibinfo {year} {2008})}\BibitemShut {NoStop}%
\bibitem [{\citenamefont {Greif}\ \emph {et~al.}(2013)\citenamefont {Greif},
  \citenamefont {Uehlinger}, \citenamefont {Jotzu}, \citenamefont {Tarruell},\
  and\ \citenamefont {Esslinger}}]{greif_short-range_2013}%
  \BibitemOpen
  \bibfield  {author} {\bibinfo {author} {\bibfnamefont {D.}~\bibnamefont
  {Greif}}, \bibinfo {author} {\bibfnamefont {T.}~\bibnamefont {Uehlinger}},
  \bibinfo {author} {\bibfnamefont {G.}~\bibnamefont {Jotzu}}, \bibinfo
  {author} {\bibfnamefont {L.}~\bibnamefont {Tarruell}}, \ and\ \bibinfo
  {author} {\bibfnamefont {T.}~\bibnamefont {Esslinger}},\ }\href {\doibase
  10.1126/science.1236362} {\bibfield  {journal} {\bibinfo  {journal}
  {Science}\ }\textbf {\bibinfo {volume} {340}},\ \bibinfo {pages} {1307}
  (\bibinfo {year} {2013})}\BibitemShut {NoStop}%
\bibitem [{\citenamefont {Hart}\ \emph {et~al.}(2015)\citenamefont {Hart},
  \citenamefont {Duarte}, \citenamefont {Yang}, \citenamefont {Liu},
  \citenamefont {Paiva}, \citenamefont {Khatami}, \citenamefont {Scalettar},
  \citenamefont {Trivedi}, \citenamefont {Huse},\ and\ \citenamefont
  {Hulet}}]{hart_observation_2015}%
  \BibitemOpen
  \bibfield  {author} {\bibinfo {author} {\bibfnamefont {R.~A.}\ \bibnamefont
  {Hart}}, \bibinfo {author} {\bibfnamefont {P.~M.}\ \bibnamefont {Duarte}},
  \bibinfo {author} {\bibfnamefont {T.-L.}\ \bibnamefont {Yang}}, \bibinfo
  {author} {\bibfnamefont {X.}~\bibnamefont {Liu}}, \bibinfo {author}
  {\bibfnamefont {T.}~\bibnamefont {Paiva}}, \bibinfo {author} {\bibfnamefont
  {E.}~\bibnamefont {Khatami}}, \bibinfo {author} {\bibfnamefont {R.~T.}\
  \bibnamefont {Scalettar}}, \bibinfo {author} {\bibfnamefont {N.}~\bibnamefont
  {Trivedi}}, \bibinfo {author} {\bibfnamefont {D.~A.}\ \bibnamefont {Huse}}, \
  and\ \bibinfo {author} {\bibfnamefont {R.~G.}\ \bibnamefont {Hulet}},\ }\href
  {\doibase 10.1038/nature14223} {\bibfield  {journal} {\bibinfo  {journal}
  {Nature}\ }\textbf {\bibinfo {volume} {519}},\ \bibinfo {pages} {211}
  (\bibinfo {year} {2015})}\BibitemShut {NoStop}%
\bibitem [{\citenamefont {Haller}\ \emph {et~al.}(2015)\citenamefont {Haller},
  \citenamefont {Hudson}, \citenamefont {Kelly}, \citenamefont {Cotta},
  \citenamefont {Peaudecerf}, \citenamefont {Bruce},\ and\ \citenamefont
  {Kuhr}}]{haller_single-atom_2015}%
  \BibitemOpen
  \bibfield  {author} {\bibinfo {author} {\bibfnamefont {E.}~\bibnamefont
  {Haller}}, \bibinfo {author} {\bibfnamefont {J.}~\bibnamefont {Hudson}},
  \bibinfo {author} {\bibfnamefont {A.}~\bibnamefont {Kelly}}, \bibinfo
  {author} {\bibfnamefont {D.~A.}\ \bibnamefont {Cotta}}, \bibinfo {author}
  {\bibfnamefont {B.}~\bibnamefont {Peaudecerf}}, \bibinfo {author}
  {\bibfnamefont {G.~D.}\ \bibnamefont {Bruce}}, \ and\ \bibinfo {author}
  {\bibfnamefont {S.}~\bibnamefont {Kuhr}},\ }\href {\doibase
  10.1038/nphys3403} {\bibfield  {journal} {\bibinfo  {journal} {Nat Phys}\
  }\textbf {\bibinfo {volume} {11}},\ \bibinfo {pages} {738} (\bibinfo {year}
  {2015})}\BibitemShut {NoStop}%
\bibitem [{\citenamefont {Edge}\ \emph {et~al.}(2015)\citenamefont {Edge},
  \citenamefont {Anderson}, \citenamefont {Jervis}, \citenamefont {McKay},
  \citenamefont {Day}, \citenamefont {Trotzky},\ and\ \citenamefont
  {Thywissen}}]{edge_imaging_2015}%
  \BibitemOpen
  \bibfield  {author} {\bibinfo {author} {\bibfnamefont {G.~J.~A.}\
  \bibnamefont {Edge}}, \bibinfo {author} {\bibfnamefont {R.}~\bibnamefont
  {Anderson}}, \bibinfo {author} {\bibfnamefont {D.}~\bibnamefont {Jervis}},
  \bibinfo {author} {\bibfnamefont {D.~C.}\ \bibnamefont {McKay}}, \bibinfo
  {author} {\bibfnamefont {R.}~\bibnamefont {Day}}, \bibinfo {author}
  {\bibfnamefont {S.}~\bibnamefont {Trotzky}}, \ and\ \bibinfo {author}
  {\bibfnamefont {J.~H.}\ \bibnamefont {Thywissen}},\ }\href {\doibase
  10.1103/PhysRevA.92.063406} {\bibfield  {journal} {\bibinfo  {journal} {Phys.
  Rev. A}\ }\textbf {\bibinfo {volume} {92}},\ \bibinfo {pages} {063406}
  (\bibinfo {year} {2015})}\BibitemShut {NoStop}%
\bibitem [{\citenamefont {Parsons}\ \emph {et~al.}(2016)\citenamefont
  {Parsons}, \citenamefont {Mazurenko}, \citenamefont {Chiu}, \citenamefont
  {Ji}, \citenamefont {Greif},\ and\ \citenamefont
  {Greiner}}]{parsons_site-resolved_2016}%
  \BibitemOpen
  \bibfield  {author} {\bibinfo {author} {\bibfnamefont {M.~F.}\ \bibnamefont
  {Parsons}}, \bibinfo {author} {\bibfnamefont {A.}~\bibnamefont {Mazurenko}},
  \bibinfo {author} {\bibfnamefont {C.~S.}\ \bibnamefont {Chiu}}, \bibinfo
  {author} {\bibfnamefont {G.}~\bibnamefont {Ji}}, \bibinfo {author}
  {\bibfnamefont {D.}~\bibnamefont {Greif}}, \ and\ \bibinfo {author}
  {\bibfnamefont {M.}~\bibnamefont {Greiner}},\ }\href {\doibase
  10.1126/science.aag1430} {\bibfield  {journal} {\bibinfo  {journal}
  {Science}\ }\textbf {\bibinfo {volume} {353}},\ \bibinfo {pages} {1253}
  (\bibinfo {year} {2016})}\BibitemShut {NoStop}%
\bibitem [{\citenamefont {Boll}\ \emph {et~al.}(2016)\citenamefont {Boll},
  \citenamefont {Hilker}, \citenamefont {Salomon}, \citenamefont {Omran},
  \citenamefont {Nespolo}, \citenamefont {Pollet}, \citenamefont {Bloch},\ and\
  \citenamefont {Gross}}]{boll_spin-_2016}%
  \BibitemOpen
  \bibfield  {author} {\bibinfo {author} {\bibfnamefont {M.}~\bibnamefont
  {Boll}}, \bibinfo {author} {\bibfnamefont {T.~A.}\ \bibnamefont {Hilker}},
  \bibinfo {author} {\bibfnamefont {G.}~\bibnamefont {Salomon}}, \bibinfo
  {author} {\bibfnamefont {A.}~\bibnamefont {Omran}}, \bibinfo {author}
  {\bibfnamefont {J.}~\bibnamefont {Nespolo}}, \bibinfo {author} {\bibfnamefont
  {L.}~\bibnamefont {Pollet}}, \bibinfo {author} {\bibfnamefont
  {I.}~\bibnamefont {Bloch}}, \ and\ \bibinfo {author} {\bibfnamefont
  {C.}~\bibnamefont {Gross}},\ }\href {\doibase 10.1126/science.aag1635}
  {\bibfield  {journal} {\bibinfo  {journal} {Science}\ }\textbf {\bibinfo
  {volume} {353}},\ \bibinfo {pages} {1257} (\bibinfo {year}
  {2016})}\BibitemShut {NoStop}%
\bibitem [{\citenamefont {Cheuk}\ \emph {et~al.}(2016)\citenamefont {Cheuk},
  \citenamefont {Nichols}, \citenamefont {Lawrence}, \citenamefont {Okan},
  \citenamefont {Zhang}, \citenamefont {Khatami}, \citenamefont {Trivedi},
  \citenamefont {Paiva}, \citenamefont {Rigol},\ and\ \citenamefont
  {Zwierlein}}]{cheuk_observation_2016}%
  \BibitemOpen
  \bibfield  {author} {\bibinfo {author} {\bibfnamefont {L.~W.}\ \bibnamefont
  {Cheuk}}, \bibinfo {author} {\bibfnamefont {M.~A.}\ \bibnamefont {Nichols}},
  \bibinfo {author} {\bibfnamefont {K.~R.}\ \bibnamefont {Lawrence}}, \bibinfo
  {author} {\bibfnamefont {M.}~\bibnamefont {Okan}}, \bibinfo {author}
  {\bibfnamefont {H.}~\bibnamefont {Zhang}}, \bibinfo {author} {\bibfnamefont
  {E.}~\bibnamefont {Khatami}}, \bibinfo {author} {\bibfnamefont
  {N.}~\bibnamefont {Trivedi}}, \bibinfo {author} {\bibfnamefont
  {T.}~\bibnamefont {Paiva}}, \bibinfo {author} {\bibfnamefont
  {M.}~\bibnamefont {Rigol}}, \ and\ \bibinfo {author} {\bibfnamefont {M.~W.}\
  \bibnamefont {Zwierlein}},\ }\href {\doibase 10.1126/science.aag3349}
  {\bibfield  {journal} {\bibinfo  {journal} {Science}\ }\textbf {\bibinfo
  {volume} {353}},\ \bibinfo {pages} {1260} (\bibinfo {year}
  {2016})}\BibitemShut {NoStop}%
\bibitem [{\citenamefont {Brown}\ \emph {et~al.}(2016)\citenamefont {Brown},
  \citenamefont {Mitra}, \citenamefont {Guardado-Sanchez}, \citenamefont
  {Schau{\ss}}, \citenamefont {Kondov}, \citenamefont {Khatami}, \citenamefont
  {Paiva}, \citenamefont {Trivedi}, \citenamefont {Huse},\ and\ \citenamefont
  {Bakr}}]{brown_observation_2016}%
  \BibitemOpen
  \bibfield  {author} {\bibinfo {author} {\bibfnamefont {P.~T.}\ \bibnamefont
  {Brown}}, \bibinfo {author} {\bibfnamefont {D.}~\bibnamefont {Mitra}},
  \bibinfo {author} {\bibfnamefont {E.}~\bibnamefont {Guardado-Sanchez}},
  \bibinfo {author} {\bibfnamefont {P.}~\bibnamefont {Schau{\ss}}}, \bibinfo
  {author} {\bibfnamefont {S.~S.}\ \bibnamefont {Kondov}}, \bibinfo {author}
  {\bibfnamefont {E.}~\bibnamefont {Khatami}}, \bibinfo {author} {\bibfnamefont
  {T.}~\bibnamefont {Paiva}}, \bibinfo {author} {\bibfnamefont
  {N.}~\bibnamefont {Trivedi}}, \bibinfo {author} {\bibfnamefont {D.~A.}\
  \bibnamefont {Huse}}, \ and\ \bibinfo {author} {\bibfnamefont {W.~S.}\
  \bibnamefont {Bakr}},\ }\href {https://arxiv.org/abs/1612.07746} {\
  (\bibinfo {year} {2016})}\BibitemShut {NoStop}%
\bibitem [{\citenamefont {Starykh}(2015)}]{starykh_unusual_2015}%
  \BibitemOpen
  \bibfield  {author} {\bibinfo {author} {\bibfnamefont {O.~A.}\ \bibnamefont
  {Starykh}},\ }\href {\doibase 10.1088/0034-4885/78/5/052502} {\bibfield
  {journal} {\bibinfo  {journal} {Rep. Prog. Phys.}\ }\textbf {\bibinfo
  {volume} {78}},\ \bibinfo {pages} {052502} (\bibinfo {year}
  {2015})}\BibitemShut {NoStop}%
\bibitem [{\citenamefont {Zhou}\ \emph {et~al.}(2017)\citenamefont {Zhou},
  \citenamefont {Kanoda},\ and\ \citenamefont {Ng}}]{zhou_quantum_2017}%
  \BibitemOpen
  \bibfield  {author} {\bibinfo {author} {\bibfnamefont {Y.}~\bibnamefont
  {Zhou}}, \bibinfo {author} {\bibfnamefont {K.}~\bibnamefont {Kanoda}}, \ and\
  \bibinfo {author} {\bibfnamefont {T.-K.}\ \bibnamefont {Ng}},\ }\href
  {\doibase 10.1103/RevModPhys.89.025003} {\bibfield  {journal} {\bibinfo
  {journal} {Rev. Mod. Phys.}\ }\textbf {\bibinfo {volume} {89}},\ \bibinfo
  {pages} {025003} (\bibinfo {year} {2017})}\BibitemShut {NoStop}%
\bibitem [{\citenamefont {Lee}\ \emph {et~al.}(2006)\citenamefont {Lee},
  \citenamefont {Nagaosa},\ and\ \citenamefont {Wen}}]{lee_doping_2006}%
  \BibitemOpen
  \bibfield  {author} {\bibinfo {author} {\bibfnamefont {P.~A.}\ \bibnamefont
  {Lee}}, \bibinfo {author} {\bibfnamefont {N.}~\bibnamefont {Nagaosa}}, \ and\
  \bibinfo {author} {\bibfnamefont {X.-G.}\ \bibnamefont {Wen}},\ }\href
  {\doibase 10.1103/RevModPhys.78.17} {\bibfield  {journal} {\bibinfo
  {journal} {Rev. Mod. Phys.}\ }\textbf {\bibinfo {volume} {78}},\ \bibinfo
  {pages} {17} (\bibinfo {year} {2006})}\BibitemShut {NoStop}%
\bibitem [{\citenamefont {Maier}\ \emph {et~al.}(2000)\citenamefont {Maier},
  \citenamefont {Jarrell}, \citenamefont {Pruschke},\ and\ \citenamefont
  {Keller}}]{maier_$mathitd$_2000}%
  \BibitemOpen
  \bibfield  {author} {\bibinfo {author} {\bibfnamefont {T.}~\bibnamefont
  {Maier}}, \bibinfo {author} {\bibfnamefont {M.}~\bibnamefont {Jarrell}},
  \bibinfo {author} {\bibfnamefont {T.}~\bibnamefont {Pruschke}}, \ and\
  \bibinfo {author} {\bibfnamefont {J.}~\bibnamefont {Keller}},\ }\href
  {\doibase 10.1103/PhysRevLett.85.1524} {\bibfield  {journal} {\bibinfo
  {journal} {Phys. Rev. Lett.}\ }\textbf {\bibinfo {volume} {85}},\ \bibinfo
  {pages} {1524} (\bibinfo {year} {2000})}\BibitemShut {NoStop}%
\bibitem [{\citenamefont {Lieb}\ and\ \citenamefont
  {Wu}(1968)}]{lieb_absence_1968}%
  \BibitemOpen
  \bibfield  {author} {\bibinfo {author} {\bibfnamefont {E.~H.}\ \bibnamefont
  {Lieb}}\ and\ \bibinfo {author} {\bibfnamefont {F.~Y.}\ \bibnamefont {Wu}},\
  }\href {\doibase 10.1103/PhysRevLett.20.1445} {\bibfield  {journal} {\bibinfo
   {journal} {Phys. Rev. Lett.}\ }\textbf {\bibinfo {volume} {20}},\ \bibinfo
  {pages} {1445} (\bibinfo {year} {1968})}\BibitemShut {NoStop}%
\bibitem [{\citenamefont {Staudt}\ \emph {et~al.}(2000)\citenamefont {Staudt},
  \citenamefont {Dzierzawa},\ and\ \citenamefont
  {Muramatsu}}]{staudt_phase_2000}%
  \BibitemOpen
  \bibfield  {author} {\bibinfo {author} {\bibfnamefont {R.}~\bibnamefont
  {Staudt}}, \bibinfo {author} {\bibfnamefont {M.}~\bibnamefont {Dzierzawa}}, \
  and\ \bibinfo {author} {\bibfnamefont {A.}~\bibnamefont {Muramatsu}},\ }\href
  {\doibase 10.1007/s100510070120} {\bibfield  {journal} {\bibinfo  {journal}
  {Eur. Phys. J. B}\ }\textbf {\bibinfo {volume} {17}},\ \bibinfo {pages} {411}
  (\bibinfo {year} {2000})}\BibitemShut {NoStop}%
\bibitem [{\citenamefont {Otsuka}\ \emph {et~al.}(2016)\citenamefont {Otsuka},
  \citenamefont {Yunoki},\ and\ \citenamefont
  {Sorella}}]{otsuka_universal_2016}%
  \BibitemOpen
  \bibfield  {author} {\bibinfo {author} {\bibfnamefont {Y.}~\bibnamefont
  {Otsuka}}, \bibinfo {author} {\bibfnamefont {S.}~\bibnamefont {Yunoki}}, \
  and\ \bibinfo {author} {\bibfnamefont {S.}~\bibnamefont {Sorella}},\ }\href
  {\doibase 10.1103/PhysRevX.6.011029} {\bibfield  {journal} {\bibinfo
  {journal} {Phys. Rev. X}\ }\textbf {\bibinfo {volume} {6}},\ \bibinfo {pages}
  {011029} (\bibinfo {year} {2016})}\BibitemShut {NoStop}%
\bibitem [{\citenamefont {Kantian}\ \emph {et~al.}(2016)\citenamefont
  {Kantian}, \citenamefont {Langer},\ and\ \citenamefont
  {Daley}}]{kantian_dynamical_2016}%
  \BibitemOpen
  \bibfield  {author} {\bibinfo {author} {\bibfnamefont {A.}~\bibnamefont
  {Kantian}}, \bibinfo {author} {\bibfnamefont {S.}~\bibnamefont {Langer}}, \
  and\ \bibinfo {author} {\bibfnamefont {A.~J.}\ \bibnamefont {Daley}},\ }\href
  {http://arxiv.org/abs/1609.03579} {\bibfield  {journal} {\bibinfo  {journal}
  {arXiv:1609.03579 [cond-mat]}\ } (\bibinfo {year} {2016})},\ \bibinfo {note}
  {arXiv: 1609.03579}\BibitemShut {NoStop}%
\bibitem [{\citenamefont {McKay}\ and\ \citenamefont
  {DeMarco}(2011)}]{mckay_cooling_2011}%
  \BibitemOpen
  \bibfield  {author} {\bibinfo {author} {\bibfnamefont {D.~C.}\ \bibnamefont
  {McKay}}\ and\ \bibinfo {author} {\bibfnamefont {B.}~\bibnamefont
  {DeMarco}},\ }\href {\doibase 10.1088/0034-4885/74/5/054401} {\bibfield
  {journal} {\bibinfo  {journal} {Rep. Prog. Phys.}\ }\textbf {\bibinfo
  {volume} {74}},\ \bibinfo {pages} {054401} (\bibinfo {year}
  {2011})}\BibitemShut {NoStop}%
\bibitem [{\citenamefont {Onofrio}(2017)}]{onofrio_cooling_2017}%
  \BibitemOpen
  \bibfield  {author} {\bibinfo {author} {\bibfnamefont {R.}~\bibnamefont
  {Onofrio}},\ }\href {\doibase 10.3367/UFNe.2016.07.037873} {\bibfield
  {journal} {\bibinfo  {journal} {Phys.-Usp.}\ }\textbf {\bibinfo {volume}
  {59}},\ \bibinfo {pages} {1129} (\bibinfo {year} {2017})}\BibitemShut
  {NoStop}%
\bibitem [{\citenamefont {Masuhara}\ \emph {et~al.}(1988)\citenamefont
  {Masuhara}, \citenamefont {Doyle}, \citenamefont {Sandberg}, \citenamefont
  {Kleppner}, \citenamefont {Greytak}, \citenamefont {Hess},\ and\
  \citenamefont {Kochanski}}]{masuhara_evaporative_1988}%
  \BibitemOpen
  \bibfield  {author} {\bibinfo {author} {\bibfnamefont {N.}~\bibnamefont
  {Masuhara}}, \bibinfo {author} {\bibfnamefont {J.~M.}\ \bibnamefont {Doyle}},
  \bibinfo {author} {\bibfnamefont {J.~C.}\ \bibnamefont {Sandberg}}, \bibinfo
  {author} {\bibfnamefont {D.}~\bibnamefont {Kleppner}}, \bibinfo {author}
  {\bibfnamefont {T.~J.}\ \bibnamefont {Greytak}}, \bibinfo {author}
  {\bibfnamefont {H.~F.}\ \bibnamefont {Hess}}, \ and\ \bibinfo {author}
  {\bibfnamefont {G.~P.}\ \bibnamefont {Kochanski}},\ }\href {\doibase
  10.1103/PhysRevLett.61.935} {\bibfield  {journal} {\bibinfo  {journal} {Phys.
  Rev. Lett.}\ }\textbf {\bibinfo {volume} {61}},\ \bibinfo {pages} {935}
  (\bibinfo {year} {1988})}\BibitemShut {NoStop}%
\bibitem [{\citenamefont {Davis}\ \emph {et~al.}(1995)\citenamefont {Davis},
  \citenamefont {Mewes}, \citenamefont {Andrews}, \citenamefont {van Druten},
  \citenamefont {Durfee}, \citenamefont {Kurn},\ and\ \citenamefont
  {Ketterle}}]{davis_bose-einstein_1995}%
  \BibitemOpen
  \bibfield  {author} {\bibinfo {author} {\bibfnamefont {K.~B.}\ \bibnamefont
  {Davis}}, \bibinfo {author} {\bibfnamefont {M.~O.}\ \bibnamefont {Mewes}},
  \bibinfo {author} {\bibfnamefont {M.~R.}\ \bibnamefont {Andrews}}, \bibinfo
  {author} {\bibfnamefont {N.~J.}\ \bibnamefont {van Druten}}, \bibinfo
  {author} {\bibfnamefont {D.~S.}\ \bibnamefont {Durfee}}, \bibinfo {author}
  {\bibfnamefont {D.~M.}\ \bibnamefont {Kurn}}, \ and\ \bibinfo {author}
  {\bibfnamefont {W.}~\bibnamefont {Ketterle}},\ }\href {\doibase
  10.1103/PhysRevLett.75.3969} {\bibfield  {journal} {\bibinfo  {journal}
  {Phys. Rev. Lett.}\ }\textbf {\bibinfo {volume} {75}},\ \bibinfo {pages}
  {3969} (\bibinfo {year} {1995})}\BibitemShut {NoStop}%
\bibitem [{\citenamefont {Modugno}\ \emph {et~al.}(2001)\citenamefont
  {Modugno}, \citenamefont {Ferrari}, \citenamefont {Roati}, \citenamefont
  {Brecha}, \citenamefont {Simoni},\ and\ \citenamefont
  {Inguscio}}]{modugno_bose-einstein_2001}%
  \BibitemOpen
  \bibfield  {author} {\bibinfo {author} {\bibfnamefont {G.}~\bibnamefont
  {Modugno}}, \bibinfo {author} {\bibfnamefont {G.}~\bibnamefont {Ferrari}},
  \bibinfo {author} {\bibfnamefont {G.}~\bibnamefont {Roati}}, \bibinfo
  {author} {\bibfnamefont {R.~J.}\ \bibnamefont {Brecha}}, \bibinfo {author}
  {\bibfnamefont {A.}~\bibnamefont {Simoni}}, \ and\ \bibinfo {author}
  {\bibfnamefont {M.}~\bibnamefont {Inguscio}},\ }\href {\doibase
  10.1126/science.1066687} {\bibfield  {journal} {\bibinfo  {journal}
  {Science}\ }\textbf {\bibinfo {volume} {294}},\ \bibinfo {pages} {1320}
  (\bibinfo {year} {2001})}\BibitemShut {NoStop}%
\bibitem [{\citenamefont {G{\"u}nter}\ \emph {et~al.}(2006)\citenamefont
  {G{\"u}nter}, \citenamefont {St{\"o}ferle}, \citenamefont {Moritz},
  \citenamefont {K{\"o}hl},\ and\ \citenamefont
  {Esslinger}}]{gunter_bose-fermi_2006}%
  \BibitemOpen
  \bibfield  {author} {\bibinfo {author} {\bibfnamefont {K.}~\bibnamefont
  {G{\"u}nter}}, \bibinfo {author} {\bibfnamefont {T.}~\bibnamefont
  {St{\"o}ferle}}, \bibinfo {author} {\bibfnamefont {H.}~\bibnamefont
  {Moritz}}, \bibinfo {author} {\bibfnamefont {M.}~\bibnamefont {K{\"o}hl}}, \
  and\ \bibinfo {author} {\bibfnamefont {T.}~\bibnamefont {Esslinger}},\ }\href
  {\doibase 10.1103/PhysRevLett.96.180402} {\bibfield  {journal} {\bibinfo
  {journal} {Phys. Rev. Lett.}\ }\textbf {\bibinfo {volume} {96}},\ \bibinfo
  {pages} {180402} (\bibinfo {year} {2006})}\BibitemShut {NoStop}%
\bibitem [{\citenamefont {Bernier}\ \emph {et~al.}(2009)\citenamefont
  {Bernier}, \citenamefont {Kollath}, \citenamefont {Georges}, \citenamefont
  {De~Leo}, \citenamefont {Gerbier}, \citenamefont {Salomon},\ and\
  \citenamefont {K{\"o}hl}}]{bernier_cooling_2009}%
  \BibitemOpen
  \bibfield  {author} {\bibinfo {author} {\bibfnamefont {J.-S.}\ \bibnamefont
  {Bernier}}, \bibinfo {author} {\bibfnamefont {C.}~\bibnamefont {Kollath}},
  \bibinfo {author} {\bibfnamefont {A.}~\bibnamefont {Georges}}, \bibinfo
  {author} {\bibfnamefont {L.}~\bibnamefont {De~Leo}}, \bibinfo {author}
  {\bibfnamefont {F.}~\bibnamefont {Gerbier}}, \bibinfo {author} {\bibfnamefont
  {C.}~\bibnamefont {Salomon}}, \ and\ \bibinfo {author} {\bibfnamefont
  {M.}~\bibnamefont {K{\"o}hl}},\ }\href {\doibase 10.1103/PhysRevA.79.061601}
  {\bibfield  {journal} {\bibinfo  {journal} {Phys. Rev. A}\ }\textbf {\bibinfo
  {volume} {79}},\ \bibinfo {pages} {061601} (\bibinfo {year}
  {2009})}\BibitemShut {NoStop}%
\bibitem [{\citenamefont {Mathy}\ \emph {et~al.}(2012)\citenamefont {Mathy},
  \citenamefont {Huse},\ and\ \citenamefont {Hulet}}]{mathy_enlarging_2012}%
  \BibitemOpen
  \bibfield  {author} {\bibinfo {author} {\bibfnamefont {C.~J.~M.}\
  \bibnamefont {Mathy}}, \bibinfo {author} {\bibfnamefont {D.~A.}\ \bibnamefont
  {Huse}}, \ and\ \bibinfo {author} {\bibfnamefont {R.~G.}\ \bibnamefont
  {Hulet}},\ }\href {\doibase 10.1103/PhysRevA.86.023606} {\bibfield  {journal}
  {\bibinfo  {journal} {Phys. Rev. A}\ }\textbf {\bibinfo {volume} {86}},\
  \bibinfo {pages} {023606} (\bibinfo {year} {2012})}\BibitemShut {NoStop}%
\bibitem [{\citenamefont {Noack}\ \emph {et~al.}(1994)\citenamefont {Noack},
  \citenamefont {White},\ and\ \citenamefont
  {Scalapino}}]{noack_correlations_1994}%
  \BibitemOpen
  \bibfield  {author} {\bibinfo {author} {\bibfnamefont {R.~M.}\ \bibnamefont
  {Noack}}, \bibinfo {author} {\bibfnamefont {S.~R.}\ \bibnamefont {White}}, \
  and\ \bibinfo {author} {\bibfnamefont {D.~J.}\ \bibnamefont {Scalapino}},\
  }\href {\doibase 10.1103/PhysRevLett.73.882} {\bibfield  {journal} {\bibinfo
  {journal} {Phys. Rev. Lett.}\ }\textbf {\bibinfo {volume} {73}},\ \bibinfo
  {pages} {882} (\bibinfo {year} {1994})}\BibitemShut {NoStop}%
\bibitem [{\citenamefont {Kuroki}\ \emph {et~al.}(1996)\citenamefont {Kuroki},
  \citenamefont {Kimura},\ and\ \citenamefont {Aoki}}]{kuroki_quantum_1996}%
  \BibitemOpen
  \bibfield  {author} {\bibinfo {author} {\bibfnamefont {K.}~\bibnamefont
  {Kuroki}}, \bibinfo {author} {\bibfnamefont {T.}~\bibnamefont {Kimura}}, \
  and\ \bibinfo {author} {\bibfnamefont {H.}~\bibnamefont {Aoki}},\ }\href
  {\doibase 10.1103/PhysRevB.54.R15641} {\bibfield  {journal} {\bibinfo
  {journal} {Phys. Rev. B}\ }\textbf {\bibinfo {volume} {54}},\ \bibinfo
  {pages} {R15641} (\bibinfo {year} {1996})}\BibitemShut {NoStop}%
\bibitem [{\citenamefont {Feiguin}\ and\ \citenamefont
  {Heidrich-Meisner}(2009)}]{feiguin_pair_2009}%
  \BibitemOpen
  \bibfield  {author} {\bibinfo {author} {\bibfnamefont {A.~E.}\ \bibnamefont
  {Feiguin}}\ and\ \bibinfo {author} {\bibfnamefont {F.}~\bibnamefont
  {Heidrich-Meisner}},\ }\href {\doibase 10.1103/PhysRevLett.102.076403}
  {\bibfield  {journal} {\bibinfo  {journal} {Phys. Rev. Lett.}\ }\textbf
  {\bibinfo {volume} {102}},\ \bibinfo {pages} {076403} (\bibinfo {year}
  {2009})}\BibitemShut {NoStop}%
\bibitem [{\citenamefont {Yamamoto}\ \emph {et~al.}(2009)\citenamefont
  {Yamamoto}, \citenamefont {Yamashita},\ and\ \citenamefont
  {Kawakami}}]{yamamoto_trapped_2009}%
  \BibitemOpen
  \bibfield  {author} {\bibinfo {author} {\bibfnamefont {A.}~\bibnamefont
  {Yamamoto}}, \bibinfo {author} {\bibfnamefont {M.}~\bibnamefont {Yamashita}},
  \ and\ \bibinfo {author} {\bibfnamefont {N.}~\bibnamefont {Kawakami}},\
  }\href {\doibase 10.1143/JPSJ.78.124001} {\bibfield  {journal} {\bibinfo
  {journal} {J. Phys. Soc. Jpn.}\ }\textbf {\bibinfo {volume} {78}},\ \bibinfo
  {pages} {124001} (\bibinfo {year} {2009})}\BibitemShut {NoStop}%
\bibitem [{\citenamefont {Sebby-Strabley}\ \emph {et~al.}(2006)\citenamefont
  {Sebby-Strabley}, \citenamefont {Anderlini}, \citenamefont {Jessen},\ and\
  \citenamefont {Porto}}]{sebby-strabley_lattice_2006}%
  \BibitemOpen
  \bibfield  {author} {\bibinfo {author} {\bibfnamefont {J.}~\bibnamefont
  {Sebby-Strabley}}, \bibinfo {author} {\bibfnamefont {M.}~\bibnamefont
  {Anderlini}}, \bibinfo {author} {\bibfnamefont {P.~S.}\ \bibnamefont
  {Jessen}}, \ and\ \bibinfo {author} {\bibfnamefont {J.~V.}\ \bibnamefont
  {Porto}},\ }\href {\doibase 10.1103/PhysRevA.73.033605} {\bibfield  {journal}
  {\bibinfo  {journal} {Phys. Rev. A}\ }\textbf {\bibinfo {volume} {73}},\
  \bibinfo {pages} {033605} (\bibinfo {year} {2006})}\BibitemShut {NoStop}%
\bibitem [{\citenamefont {F{\"o}lling}\ \emph {et~al.}(2007)\citenamefont
  {F{\"o}lling}, \citenamefont {Trotzky}, \citenamefont {Cheinet},
  \citenamefont {Feld}, \citenamefont {Saers}, \citenamefont {Widera},
  \citenamefont {M{\"u}ller},\ and\ \citenamefont
  {Bloch}}]{folling_direct_2007}%
  \BibitemOpen
  \bibfield  {author} {\bibinfo {author} {\bibfnamefont {S.}~\bibnamefont
  {F{\"o}lling}}, \bibinfo {author} {\bibfnamefont {S.}~\bibnamefont
  {Trotzky}}, \bibinfo {author} {\bibfnamefont {P.}~\bibnamefont {Cheinet}},
  \bibinfo {author} {\bibfnamefont {M.}~\bibnamefont {Feld}}, \bibinfo {author}
  {\bibfnamefont {R.}~\bibnamefont {Saers}}, \bibinfo {author} {\bibfnamefont
  {A.}~\bibnamefont {Widera}}, \bibinfo {author} {\bibfnamefont
  {T.}~\bibnamefont {M{\"u}ller}}, \ and\ \bibinfo {author} {\bibfnamefont
  {I.}~\bibnamefont {Bloch}},\ }\href {\doibase 10.1038/nature06112} {\bibfield
   {journal} {\bibinfo  {journal} {Nature}\ }\textbf {\bibinfo {volume}
  {448}},\ \bibinfo {pages} {1029} (\bibinfo {year} {2007})}\BibitemShut
  {NoStop}%
\bibitem [{\citenamefont {Danshita}\ \emph {et~al.}(2007)\citenamefont
  {Danshita}, \citenamefont {Williams}, \citenamefont {S{\'a}~de Melo},\ and\
  \citenamefont {Clark}}]{danshita_quantum_2007}%
  \BibitemOpen
  \bibfield  {author} {\bibinfo {author} {\bibfnamefont {I.}~\bibnamefont
  {Danshita}}, \bibinfo {author} {\bibfnamefont {J.~E.}\ \bibnamefont
  {Williams}}, \bibinfo {author} {\bibfnamefont {C.~A.~R.}\ \bibnamefont
  {S{\'a}~de Melo}}, \ and\ \bibinfo {author} {\bibfnamefont {C.~W.}\
  \bibnamefont {Clark}},\ }\href {\doibase 10.1103/PhysRevA.76.043606}
  {\bibfield  {journal} {\bibinfo  {journal} {Phys. Rev. A}\ }\textbf {\bibinfo
  {volume} {76}},\ \bibinfo {pages} {043606} (\bibinfo {year}
  {2007})}\BibitemShut {NoStop}%
\bibitem [{\citenamefont {Chen}\ \emph {et~al.}(2011)\citenamefont {Chen},
  \citenamefont {Huber}, \citenamefont {Trotzky}, \citenamefont {Bloch},\ and\
  \citenamefont {Altman}}]{chen_many-body_2011}%
  \BibitemOpen
  \bibfield  {author} {\bibinfo {author} {\bibfnamefont {Y.-A.}\ \bibnamefont
  {Chen}}, \bibinfo {author} {\bibfnamefont {S.~D.}\ \bibnamefont {Huber}},
  \bibinfo {author} {\bibfnamefont {S.}~\bibnamefont {Trotzky}}, \bibinfo
  {author} {\bibfnamefont {I.}~\bibnamefont {Bloch}}, \ and\ \bibinfo {author}
  {\bibfnamefont {E.}~\bibnamefont {Altman}},\ }\href {\doibase
  10.1038/nphys1801} {\bibfield  {journal} {\bibinfo  {journal} {Nat Phys}\
  }\textbf {\bibinfo {volume} {7}},\ \bibinfo {pages} {61} (\bibinfo {year}
  {2011})}\BibitemShut {NoStop}%
\bibitem [{\citenamefont
  {Schollw{\"o}ck}(2011)}]{schollwock_density-matrix_2011}%
  \BibitemOpen
  \bibfield  {author} {\bibinfo {author} {\bibfnamefont {U.}~\bibnamefont
  {Schollw{\"o}ck}},\ }\href {\doibase 10.1016/j.aop.2010.09.012} {\bibfield
  {journal} {\bibinfo  {journal} {Annals of Physics}\ }\textbf {\bibinfo
  {volume} {326}},\ \bibinfo {pages} {96} (\bibinfo {year} {2011})}\BibitemShut
  {NoStop}%
\bibitem [{\citenamefont {White}(2009)}]{white_minimally_2009}%
  \BibitemOpen
  \bibfield  {author} {\bibinfo {author} {\bibfnamefont {S.~R.}\ \bibnamefont
  {White}},\ }\href {\doibase 10.1103/PhysRevLett.102.190601} {\bibfield
  {journal} {\bibinfo  {journal} {Phys. Rev. Lett.}\ }\textbf {\bibinfo
  {volume} {102}},\ \bibinfo {pages} {190601} (\bibinfo {year}
  {2009})}\BibitemShut {NoStop}%
\bibitem [{\citenamefont {Stoudenmire}\ and\ \citenamefont
  {White}(2010)}]{stoudenmire_minimally_2010}%
  \BibitemOpen
  \bibfield  {author} {\bibinfo {author} {\bibfnamefont {E.~M.}\ \bibnamefont
  {Stoudenmire}}\ and\ \bibinfo {author} {\bibfnamefont {S.~R.}\ \bibnamefont
  {White}},\ }\href {\doibase 10.1088/1367-2630/12/5/055026} {\bibfield
  {journal} {\bibinfo  {journal} {New J. Phys.}\ }\textbf {\bibinfo {volume}
  {12}},\ \bibinfo {pages} {055026} (\bibinfo {year} {2010})}\BibitemShut
  {NoStop}%
\bibitem [{\citenamefont {Verstraete}\ \emph {et~al.}(2004)\citenamefont
  {Verstraete}, \citenamefont {Garc{\'i}a-Ripoll},\ and\ \citenamefont
  {Cirac}}]{verstraete_matrix_2004}%
  \BibitemOpen
  \bibfield  {author} {\bibinfo {author} {\bibfnamefont {F.}~\bibnamefont
  {Verstraete}}, \bibinfo {author} {\bibfnamefont {J.~J.}\ \bibnamefont
  {Garc{\'i}a-Ripoll}}, \ and\ \bibinfo {author} {\bibfnamefont {J.~I.}\
  \bibnamefont {Cirac}},\ }\href {\doibase 10.1103/PhysRevLett.93.207204}
  {\bibfield  {journal} {\bibinfo  {journal} {Phys. Rev. Lett.}\ }\textbf
  {\bibinfo {volume} {93}},\ \bibinfo {pages} {207204} (\bibinfo {year}
  {2004})}\BibitemShut {NoStop}%
\bibitem [{\citenamefont {Feiguin}\ and\ \citenamefont
  {White}(2005)}]{feiguin_finite-temperature_2005}%
  \BibitemOpen
  \bibfield  {author} {\bibinfo {author} {\bibfnamefont {A.~E.}\ \bibnamefont
  {Feiguin}}\ and\ \bibinfo {author} {\bibfnamefont {S.~R.}\ \bibnamefont
  {White}},\ }\href {\doibase 10.1103/PhysRevB.72.220401} {\bibfield  {journal}
  {\bibinfo  {journal} {Phys. Rev. B}\ }\textbf {\bibinfo {volume} {72}},\
  \bibinfo {pages} {220401} (\bibinfo {year} {2005})}\BibitemShut {NoStop}%
\bibitem [{\citenamefont {Binder}\ and\ \citenamefont
  {Barthel}(2015)}]{binder_minimally_2015}%
  \BibitemOpen
  \bibfield  {author} {\bibinfo {author} {\bibfnamefont {M.}~\bibnamefont
  {Binder}}\ and\ \bibinfo {author} {\bibfnamefont {T.}~\bibnamefont
  {Barthel}},\ }\href {\doibase 10.1103/PhysRevB.92.125119} {\bibfield
  {journal} {\bibinfo  {journal} {Phys. Rev. B}\ }\textbf {\bibinfo {volume}
  {92}},\ \bibinfo {pages} {125119} (\bibinfo {year} {2015})}\BibitemShut
  {NoStop}%
\bibitem [{\citenamefont {Barthel}(2016)}]{barthel_matrix_2016}%
  \BibitemOpen
  \bibfield  {author} {\bibinfo {author} {\bibfnamefont {T.}~\bibnamefont
  {Barthel}},\ }\href {\doibase 10.1103/PhysRevB.94.115157} {\bibfield
  {journal} {\bibinfo  {journal} {Phys. Rev. B}\ }\textbf {\bibinfo {volume}
  {94}},\ \bibinfo {pages} {115157} (\bibinfo {year} {2016})}\BibitemShut
  {NoStop}%
\bibitem [{\citenamefont {Karrasch}\ \emph {et~al.}(2012)\citenamefont
  {Karrasch}, \citenamefont {Bardarson},\ and\ \citenamefont
  {Moore}}]{karrasch_finite-temperature_2012}%
  \BibitemOpen
  \bibfield  {author} {\bibinfo {author} {\bibfnamefont {C.}~\bibnamefont
  {Karrasch}}, \bibinfo {author} {\bibfnamefont {J.~H.}\ \bibnamefont
  {Bardarson}}, \ and\ \bibinfo {author} {\bibfnamefont {J.~E.}\ \bibnamefont
  {Moore}},\ }\href {\doibase 10.1103/PhysRevLett.108.227206} {\bibfield
  {journal} {\bibinfo  {journal} {Phys. Rev. Lett.}\ }\textbf {\bibinfo
  {volume} {108}},\ \bibinfo {pages} {227206} (\bibinfo {year}
  {2012})}\BibitemShut {NoStop}%
\bibitem [{\citenamefont {Karrasch}\ \emph {et~al.}(2013)\citenamefont
  {Karrasch}, \citenamefont {Bardarson},\ and\ \citenamefont
  {Moore}}]{karrasch_reducing_2013}%
  \BibitemOpen
  \bibfield  {author} {\bibinfo {author} {\bibfnamefont {C.}~\bibnamefont
  {Karrasch}}, \bibinfo {author} {\bibfnamefont {J.~H.}\ \bibnamefont
  {Bardarson}}, \ and\ \bibinfo {author} {\bibfnamefont {J.~E.}\ \bibnamefont
  {Moore}},\ }\href {\doibase 10.1088/1367-2630/15/8/083031} {\bibfield
  {journal} {\bibinfo  {journal} {New J. Phys.}\ }\textbf {\bibinfo {volume}
  {15}},\ \bibinfo {pages} {083031} (\bibinfo {year} {2013})}\BibitemShut
  {NoStop}%
\bibitem [{\citenamefont {Tiegel}\ \emph {et~al.}(2014)\citenamefont {Tiegel},
  \citenamefont {Manmana}, \citenamefont {Pruschke},\ and\ \citenamefont
  {Honecker}}]{tiegel_matrix_2014}%
  \BibitemOpen
  \bibfield  {author} {\bibinfo {author} {\bibfnamefont {A.~C.}\ \bibnamefont
  {Tiegel}}, \bibinfo {author} {\bibfnamefont {S.~R.}\ \bibnamefont {Manmana}},
  \bibinfo {author} {\bibfnamefont {T.}~\bibnamefont {Pruschke}}, \ and\
  \bibinfo {author} {\bibfnamefont {A.}~\bibnamefont {Honecker}},\ }\href
  {\doibase 10.1103/PhysRevB.90.060406} {\bibfield  {journal} {\bibinfo
  {journal} {Phys. Rev. B}\ }\textbf {\bibinfo {volume} {90}},\ \bibinfo
  {pages} {060406} (\bibinfo {year} {2014})}\BibitemShut {NoStop}%
\bibitem [{\citenamefont {Bruognolo}\ \emph {et~al.}(2017)\citenamefont
  {Bruognolo}, \citenamefont {Zhu}, \citenamefont {White},\ and\ \citenamefont
  {Stoudenmire}}]{bruognolo_matrix_2017}%
  \BibitemOpen
  \bibfield  {author} {\bibinfo {author} {\bibfnamefont {B.}~\bibnamefont
  {Bruognolo}}, \bibinfo {author} {\bibfnamefont {Z.}~\bibnamefont {Zhu}},
  \bibinfo {author} {\bibfnamefont {S.~R.}\ \bibnamefont {White}}, \ and\
  \bibinfo {author} {\bibfnamefont {E.~M.}\ \bibnamefont {Stoudenmire}},\
  }\href {http://arxiv.org/abs/1705.05578} {\bibfield  {journal} {\bibinfo
  {journal} {arXiv:1705.05578 [cond-mat]}\ } (\bibinfo {year} {2017})},\
  \bibinfo {note} {arXiv: 1705.05578}\BibitemShut {NoStop}%
\bibitem [{\citenamefont {Brown}\ \emph {et~al.}(1989)\citenamefont {Brown},
  \citenamefont {Byrne},\ and\ \citenamefont {Hindmarsh}}]{brown_vode:_1989}%
  \BibitemOpen
  \bibfield  {author} {\bibinfo {author} {\bibfnamefont {P.}~\bibnamefont
  {Brown}}, \bibinfo {author} {\bibfnamefont {G.}~\bibnamefont {Byrne}}, \ and\
  \bibinfo {author} {\bibfnamefont {A.}~\bibnamefont {Hindmarsh}},\ }\href
  {\doibase 10.1137/0910062} {\bibfield  {journal} {\bibinfo  {journal} {SIAM
  J. Sci. and Stat. Comput.}\ }\textbf {\bibinfo {volume} {10}},\ \bibinfo
  {pages} {1038} (\bibinfo {year} {1989})}\BibitemShut {NoStop}%
\bibitem [{\citenamefont {Fedichev}\ \emph {et~al.}(1996)\citenamefont
  {Fedichev}, \citenamefont {Kagan}, \citenamefont {Shlyapnikov},\ and\
  \citenamefont {Walraven}}]{fedichev_influence_1996}%
  \BibitemOpen
  \bibfield  {author} {\bibinfo {author} {\bibfnamefont {P.~O.}\ \bibnamefont
  {Fedichev}}, \bibinfo {author} {\bibfnamefont {Y.}~\bibnamefont {Kagan}},
  \bibinfo {author} {\bibfnamefont {G.~V.}\ \bibnamefont {Shlyapnikov}}, \ and\
  \bibinfo {author} {\bibfnamefont {J.~T.~M.}\ \bibnamefont {Walraven}},\
  }\href {\doibase 10.1103/PhysRevLett.77.2913} {\bibfield  {journal} {\bibinfo
   {journal} {Phys. Rev. Lett.}\ }\textbf {\bibinfo {volume} {77}},\ \bibinfo
  {pages} {2913} (\bibinfo {year} {1996})}\BibitemShut {NoStop}%
\bibitem [{\citenamefont {Yamazaki}\ \emph {et~al.}(2010)\citenamefont
  {Yamazaki}, \citenamefont {Taie}, \citenamefont {Sugawa},\ and\ \citenamefont
  {Takahashi}}]{yamazaki_submicron_2010}%
  \BibitemOpen
  \bibfield  {author} {\bibinfo {author} {\bibfnamefont {R.}~\bibnamefont
  {Yamazaki}}, \bibinfo {author} {\bibfnamefont {S.}~\bibnamefont {Taie}},
  \bibinfo {author} {\bibfnamefont {S.}~\bibnamefont {Sugawa}}, \ and\ \bibinfo
  {author} {\bibfnamefont {Y.}~\bibnamefont {Takahashi}},\ }\href {\doibase
  10.1103/PhysRevLett.105.050405} {\bibfield  {journal} {\bibinfo  {journal}
  {Phys. Rev. Lett.}\ }\textbf {\bibinfo {volume} {105}},\ \bibinfo {pages}
  {050405} (\bibinfo {year} {2010})}\BibitemShut {NoStop}%
\bibitem [{\citenamefont {Clark}\ \emph {et~al.}(2015)\citenamefont {Clark},
  \citenamefont {Ha}, \citenamefont {Xu},\ and\ \citenamefont
  {Chin}}]{clark_quantum_2015}%
  \BibitemOpen
  \bibfield  {author} {\bibinfo {author} {\bibfnamefont {L.~W.}\ \bibnamefont
  {Clark}}, \bibinfo {author} {\bibfnamefont {L.-C.}\ \bibnamefont {Ha}},
  \bibinfo {author} {\bibfnamefont {C.-Y.}\ \bibnamefont {Xu}}, \ and\ \bibinfo
  {author} {\bibfnamefont {C.}~\bibnamefont {Chin}},\ }\href {\doibase
  10.1103/PhysRevLett.115.155301} {\bibfield  {journal} {\bibinfo  {journal}
  {Phys. Rev. Lett.}\ }\textbf {\bibinfo {volume} {115}},\ \bibinfo {pages}
  {155301} (\bibinfo {year} {2015})}\BibitemShut {NoStop}%
\bibitem [{\citenamefont {Garc{\'i}a-Ripoll}(2006)}]{garcia-ripoll_time_2006}%
  \BibitemOpen
  \bibfield  {author} {\bibinfo {author} {\bibfnamefont {J.~J.}\ \bibnamefont
  {Garc{\'i}a-Ripoll}},\ }\href {\doibase 10.1088/1367-2630/8/12/305}
  {\bibfield  {journal} {\bibinfo  {journal} {New J. Phys.}\ }\textbf {\bibinfo
  {volume} {8}},\ \bibinfo {pages} {305} (\bibinfo {year} {2006})}\BibitemShut
  {NoStop}%
\bibitem [{\citenamefont {Zaletel}\ \emph {et~al.}(2015)\citenamefont
  {Zaletel}, \citenamefont {Mong}, \citenamefont {Karrasch}, \citenamefont
  {Moore},\ and\ \citenamefont {Pollmann}}]{zaletel_time-evolving_2015}%
  \BibitemOpen
  \bibfield  {author} {\bibinfo {author} {\bibfnamefont {M.~P.}\ \bibnamefont
  {Zaletel}}, \bibinfo {author} {\bibfnamefont {R.~S.~K.}\ \bibnamefont
  {Mong}}, \bibinfo {author} {\bibfnamefont {C.}~\bibnamefont {Karrasch}},
  \bibinfo {author} {\bibfnamefont {J.~E.}\ \bibnamefont {Moore}}, \ and\
  \bibinfo {author} {\bibfnamefont {F.}~\bibnamefont {Pollmann}},\ }\href
  {\doibase 10.1103/PhysRevB.91.165112} {\bibfield  {journal} {\bibinfo
  {journal} {Phys. Rev. B}\ }\textbf {\bibinfo {volume} {91}},\ \bibinfo
  {pages} {165112} (\bibinfo {year} {2015})}\BibitemShut {NoStop}%
\bibitem [{\citenamefont {Corboz}\ and\ \citenamefont
  {Vidal}(2009)}]{corboz_fermionic_2009}%
  \BibitemOpen
  \bibfield  {author} {\bibinfo {author} {\bibfnamefont {P.}~\bibnamefont
  {Corboz}}\ and\ \bibinfo {author} {\bibfnamefont {G.}~\bibnamefont {Vidal}},\
  }\href {\doibase 10.1103/PhysRevB.80.165129} {\bibfield  {journal} {\bibinfo
  {journal} {Phys. Rev. B}\ }\textbf {\bibinfo {volume} {80}},\ \bibinfo
  {pages} {165129} (\bibinfo {year} {2009})}\BibitemShut {NoStop}%
\end{thebibliography}%
%\input{main.bbl}
%\bibliography{library}

\end{document}